\documentclass[twocolumn,twocolappendix,trackchanges,
]{aastex631} %linenumbers

\usepackage{amsmath}
\usepackage{subfigure}
\usepackage{float}

\newcommand*{\Msun}{\rm{M}_\odot}
\newcommand*{\Mstar}{M_\star}
\newcommand*{\Mvir}{M_{\mathrm{vir}}}
\newcommand*{\Mpeak}{M_{\mathrm{peak}}}
\newcommand*{\Rvir}{R_{\mathrm{vir}}}
\newcommand*{\LCDM}{\Lambda\mathrm{CDM}}

\newcommand*{\Nint}{N_{\mathrm{interloper}}}
\newcommand*{\Nrand}{N_{\mathrm{random}}}
\newcommand*{\avgNhost}{\langle N_{\mathrm{host}} \rangle}
\newcommand*{\avgNrand}{\langle N_{\mathrm{random}} \rangle}
\newcommand*{\fdm}{f_{\mathrm{DM}}}
\newcommand*{\fstr}{f_\star}

\shorttitle{Satellites and subhalos at cosmic noon}
\shortauthors{J. T. Wan et al.}

\begin{document}
    
\title{Dwarf Galaxies at Cosmic Noon: \\ New JWST Constraints on Satellite Models and Subhalo Tidal Evolution}

\correspondingauthor{Jenny T. Wan}
\email{jennywan@stanford.edu}

\author{Jenny T. Wan}
\affiliation{Kavli Institute for Particle Astrophysics \& Cosmology, P.O. Box 2450, Stanford University, Stanford, CA 94305, USA}
\affiliation{Department of Physics, Stanford University, 382 Via Pueblo Mall, Stanford, CA 94305, USA}

\author{Philip Mansfield}
\affiliation{Kavli Institute for Particle Astrophysics \& Cosmology, P.O. Box 2450, Stanford University, Stanford, CA 94305, USA}

\author{Katherine A. Suess} 
\affiliation{Department for Astrophysical \& Planetary Science, University of Colorado, Boulder, CO 80309, USA}

\author{Yunchong Wang} 
\affiliation{Kavli Institute for Particle Astrophysics \& Cosmology, P.O. Box 2450, Stanford University, Stanford, CA 94305, USA}
\affiliation{Department of Physics, Stanford University, 382 Via Pueblo Mall, Stanford, CA 94305, USA}
\affiliation{SLAC National Accelerator Laboratory, Menlo Park, CA 94025, USA}

\author{Sihan Yuan} 
\affiliation{Kavli Institute for Particle Astrophysics \& Cosmology, P.O. Box 2450, Stanford University, Stanford, CA 94305, USA}
\affiliation{SLAC National Accelerator Laboratory, Menlo Park, CA 94025, USA}

\author{Christina C. Williams} 
\affiliation{NSF National Optical-Infrared Astronomy Research Laboratory, 950 North Cherry Avenue, Tucson, AZ 85719, USA}

\author{Risa H. Wechsler}
\affiliation{Kavli Institute for Particle Astrophysics \& Cosmology, P.O. Box 2450, Stanford University, Stanford, CA 94305, USA}
\affiliation{Department of Physics, Stanford University, 382 Via Pueblo Mall, Stanford, CA 94305, USA}
\affiliation{SLAC National Accelerator Laboratory, Menlo Park, CA 94025, USA}

\begin{abstract}

The advent of JWST has revolutionized the study of faint satellite galaxies at $z \gtrsim 1$, enabling statistical constraints on galaxy evolution and the galaxy--halo connection in a previously unexplored mass and redshift regime. We compare satellite abundances at $1 < z < 3.5$ from recent JWST observations with predictions from cosmological dark matter-only zoom-in simulations. We identify and quantify several sources of biases that can impact theoretical satellite counts, finding that assumptions about subhalo tidal evolution introduce the largest uncertainty in predictions for the satellite mass function. Using a flexible galaxy disruption model, we explore a range of disruption scenarios, spanning hydrodynamically motivated and idealized prescriptions, to bracket plausible physical outcomes. We show that varying galaxy durability can change the predicted satellite mass functions by a factor of $\sim3.5$. The JWST data and our fiducial model are consistent within $1-2\sigma$ across the full redshift ($1 < z < 3.5$) and stellar mass ($\Mstar > 10^7~\Msun$) range probed. We find evidence that subhalos are at least as long-lived as predicted by hydrodynamic simulations. Our framework will enable robust constraints on the tidal evolution of subhalos with future observations. This work presents the first direct comparison between cosmological models and observations of the high-redshift satellite population in this low-mass regime. These results showcase JWST's emerging power to test structure formation in the first half of the Universe in a new domain and to constrain the physical processes driving the evolution of low-mass galaxies across cosmic time.

\end{abstract}

\keywords{\href{http://astrothesaurus.org/uat/416}{Dwarf galaxies (416)}, \href{http://astrothesaurus.org/uat/594}{Galaxy evolution (594)}, \href{http://astrothesaurus.org/uat/599}{Galaxy infall (599)}, \href{http://astrothesaurus.org/uat/734}{High-redshift galaxies (734)}, \href{http://astrothesaurus.org/uat/1880}{Galaxy dark matter halos (1880)}}

%%%%%%%%%%%%%%%%%%%%%%%%%%%%%%%%%%%%%%%%%%%%%%%%%%%%%%%%%%%%%%%%%%%%%%%%

\section{Introduction} \label{sec:intro}

Satellite galaxies, the companions orbiting more massive host galaxies, are invaluable probes of a wide range of physical processes fundamental to our understanding of the Universe. Measurements of satellite abundances around the Milky Way (MW) and Andromeda (M31), which initially posed challenges to the predictions of the standard $\LCDM$ model of cosmology (e.g., the ``missing satellites'' and ``too big to fail'' problems; \citealt{Moore1999, Klypin1999, TBTF2011, TBTF2012, BBK2017}), have led to major advances in our understanding of the local satellite population and its observational completeness \cite{Drlica-Wagner2020}. In the last few years, the SAGA \citep{SAGA2017, SAGA2024} and ELVES \citep{ELVES2022} surveys have expanded studies of faint satellites to the Local Volume, and a range of studies have compared these measurements to models to understand the low-mass galaxy--halo connection \citep{Newton2018, Kim2018, MWSatCensusII, ELVES_IV} and the impact of baryonic physics on satellite distributions \citep{Benson2002, Somerville2002, Kravtsov2004a, APOSTLE2016, Wetzel2016, GarrisonKimmel2017, Lovell2017, Kelley2019, EDEN2024}. Together, these efforts have largely resolved the apparent tensions between observations and simulations of satellites in the local universe. The satellite radial distributions, quenched fractions, specific star-formation rates, luminosity functions, and stellar mass functions (SMFs) measured in the Milky Way and the Local Volume now serve as key anchor points for calibrating galaxy formation models \citep[e.g.,][]{Font2011, SAGA2017, IllustrisTNG2019, UM2019, MWSatCensusII, ELVES2022, Wu2022, ELVES_IV, SAGA2024, UM-SAGA}.

Historically, constraints based on faint systems have been limited to the Local Volume due to the difficulty of observing faint, low-mass galaxies at large cosmological distances. Measurements from the Hubble Space Telescope expanded such studies significantly \citep[e.g.,][]{Newman2012, Nierenberg2013, Kawinwanichakij2016, Roberts2021}. However, even when considering quiescent, massive ($\log_{10} \Mstar/\Msun \gtrsim 10.5$) host galaxies, measurements from HST and deep ground-based programs have only been able to robustly identify satellites beyond $z \sim 1$ with mass ratios of $m_{\rm \star,sat}/m_{\rm \star,host}\sim 1/10$. 

The advent of JWST \citep{Gardner2023} is now enabling measurements of faint companions around massive hosts at redshifts $z \gtrsim 1$ for the first time \citep{Suess2023}. These newly observable satellite galaxies, selected to lie within projected separations of $\lesssim 35$ kpc from their hosts, have a median satellite-to-host stellar mass ratio of just $1/900$, meaning we are probing a \textit{fundamentally new regime} --- statistical samples of dwarf galaxies in the inner regions of host halos at high redshift. This enables us to address several unresolved questions in galaxy evolution, including the role of gas-poor minor mergers in the structural evolution of massive galaxies \citep[e.g.,][]{Bezanson2009, Slob2025} and the origin the observed correlation between the star formation properties of central galaxies and their satellites, known as galactic conformity, thus far mostly explored at $z \lesssim 0.2$ \citep{Weinmann2006, Hearin2016}. 

Before we can use observations of the high-redshift satellite population to meaningfully test the predictions of cosmological models --- especially those that rely on galaxy baryonic properties --- it is critical to fully understand the systematic uncertainties and biases in these theoretical models. The primary goal of this work is to develop a modeling methodology that enables consistent comparisons between the predicted satellite population in the early universe and JWST observations, thereby allowing us to test and constrain models of galaxy formation and evolution.

In particular, we must first establish the translation from simulated dark matter halos to observed galaxies, a relationship known as the ``galaxy--halo connection'' \citep{WechslerTinker2018}. The galaxy--halo connection is reasonably well-constrained for galaxies with stellar masses $\log_{10} \Mstar/\Msun \gtrsim$ 9--10 at redshifts $z < 1$. However, it remains highly uncertain at lower masses or higher redshifts, and even less certain for satellites at very close separations from their central host galaxies \citep{WechslerTinker2018}. 

One major source of this uncertainty stems from our limited understanding of how satellite galaxies and the subhalos they reside in evolve after falling into the gravitational potentials of their host halos (``post-infall''). As subhalos orbit their hosts, they experience strong tidal forces that strip mass from their outskirts (``tidal stripping''). Accurately simulating these small, heavily stripped systems is a significant challenge for two key reasons. First, halo finders, the tool used in simulations to identify and track halos and subhalos across time, often lose or mischaracterize subhalos even when they retain substantial mass \citep[e.g.,][]{Symfind2024}. Second, the computational need to model the continuous dark matter field as a discrete $N$-body system introduces numerical effects that lead to artificial, premature disruption, even in relatively high-resolution subhalos \citep[e.g.,][]{ vdBosch2017, vdBosch2018, vdBoschOgiya2018}. These effects are discussed in more detail in Section \ref{sec:numerical disruption}. 

Together, halo finder errors and numerical artifacts can severely compromise our ability to trust predictions for the survivability of simulated subhalos, as well as their mass and structural evolution after infall. This is particularly problematic for low-mass satellites, as it directly impacts our understanding of the coupling between \textit{subhalo} (dark matter) and \textit{galaxy} (stellar) tidal mass loss. Recent advances in particle tracking-based halo finders such as \textsc{HBT+} \citep{HBT+2018}, \textsc{Symfind} \citep{Symfind2024}, \textsc{Sparta} \citep{Diemer2024}, and \textsc{Bloodhound} \citep{Kong2025} are now allowing for robust subhalo tracking down to low masses, but even a perfect subhalo finder would not be able to fully eliminate the impact of numerical disruption. As such, a central motivation for our work is to develop an improved model for subhalo tidal evolution that is as free as possible from numerical limitations.

In this work, we significantly improve the modeling of low-mass, heavily stripped subhalos in high-resolution, dark matter-only cosmological simulations and assess the consistency between simulated satellite abundances and JWST observations of satellite galaxies at $1 < z < 3.5$. We quantify and correct for the main sources of bias that impact theoretical satellite counts, including host distribution mismatches, halo contraction due to the host's baryons, ``interlopers'' from large-scale structure, simulation numerics, and systematic uncertainty in galaxy durability. To do this, we implement a particle-based orphan model to approximate the properties of subhalos that are no longer numerically converged, as well as a flexible, tidal track-based model for galaxy tidal mass loss to explore the effect of galaxy durability on predicted satellite abundances. We compare our resulting theoretical satellite stellar mass functions with the observed SMFs and demonstrate the potential of our model to constrain the physics of subhalo tidal evolution.

This paper is structured as follows. In Section \ref{sec:data}, we describe both the observations and simulations used throughout this work. Section \ref{sec:methods} describes the modeling steps applied to the simulated subhalo population in order to perform a one-to-one comparison to the observations. We present our results in Section \ref{sec:results} and address key sources of systematics in Section \ref{sec:systematics}. We discuss the implications of our work and conclude in Section \ref{sec:conclusions}. 

Throughout this work, we assume a $\LCDM$ cosmology with parameters $h = 0.7$, $\Omega_m = 0.286$, $\Omega_\Lambda = 0.714$, and $\sigma_8 = 0.82$.

%%%%%%%%%%%%%%%%%%%%%%%%%%%%%%%%%%%%%%%%%%%%%%%%%%%%%%%%%%%%%%%%%%%%%%%%%

\section{Data} \label{sec:data}

\subsection{Observations} \label{sec:observations}

The observational data used in this work come from the sample of hosts and satellites identified by \citet{Suess2023} using multiband JWST/NIRCam mosaics in GOODS-S, which included imaging from JADES \citep{JADES2023}, JEMS \citep{JEMS2023}, and FRESCO \citep{FRESCO2023}. We refer the reader to \citet{Suess2023} for a more detailed description of the observations, as well as source detection and characterization methods. In brief, quiescent host galaxies with redshifts $z \geq 0.5$ were selected to have $\log \Mstar/\Msun \geq 10$ and rest-frame \textit{UVJ} colors that lie within the \citet{Schreiber2015} quiescent box. Here, we focus on systems in the redshift range $1 < z < 3.5$, resulting in 93 hosts in our observational sample. Galaxies within a 35 kpc radius of each host that have photometric redshifts consistent within $1\sigma$ of the host redshift are classified as satellite galaxies. 

We assume mass-completeness limits that are 1.5 dex below the 90\% completeness limits of CANDELS-wide from \citet{Tal2014}, corresponding to the depth difference between HST/F160W imaging and JADES JWST/F150W imaging \citep{JADES2023, Suess2023}. These thresholds are $10^{7.04}~\Msun$ at $1 < z < 1.2$, $10^{7.56}~\Msun$ at $1.2 < z < 1.8$, $10^{7.91}~\Msun$ at $1.8 < z < 2.5$, and $10^{9.14}~\Msun$ at $2.5 < z < 3.5$. We list the mass-completeness limits used in this work in Table \ref{tab:observed hosts}, along with the number of observed hosts that fall within each bin. These limits are conservative, and a more detailed completeness analysis may allow one to push the data further towards a larger, deeper sample.

\begin{table}
    \centering
    \begin{tabular}{lcc}
        \hline
        \hline
        Redshift & Mass-completeness & Number of \\
        bin & threshold & hosts \\
        \hline
        $1.0 < z < 1.2$ & $10^{7.04}~\Msun$ & 21 \\ 

$1.2 < z < 1.8$ & $10^{7.56}~\Msun$ & 37 \\ 

$1.8 < z < 2.5$ & $10^{7.91}~\Msun$ & 16 \\

$2.5 < z < 3.5$ & $10^{9.14}~\Msun$ & 19 \\ \\
        \hline
    \end{tabular}    
    \caption{The redshift bins used in our analysis, along with the corresponding mass-completeness limits, and number of hosts per bin.}
    \label{tab:observed hosts}
\end{table}

\subsection{Simulations} \label{sec:simulations}

The simulations used in this work are part of Symphony, a compilation of cold dark matter-only cosmological zoom-in simulations \citep{Nadler2023}. We analyze the ``SymphonyGroup'' suite, a set of zoom-in simulations containing 49 host halos at the $\Mvir \sim 10^{13}~\Msun$ scale. The host halos for the suite were chosen from the parent simulation (c125-1024, which has a side length of 125 Mpc $h^{-1}$ and 1024 particles per side) such that their virial masses are $10^{12.86 \pm 0.10}~\Msun$ at $z = 0.5$. % This sharply peaked mass distrubtion does have consequences, which touch on a little later in this section (see also Section 3.1 and Appendix A). However, among the available Symphony zoom-in suites, SymphonyGroup contains hosts with masses most comparable to those in our observational sample.
We chose this specific suite because, among the available Symphony zoom-in suites, SymphonyGroup contains hosts with masses most comparable to those in our observational sample.

The highest resolution particles of the SymphonyGroup zoom-ins have masses of $m_{\rm{part}} = 3.3 \times 10^6~\Msun$. The comoving Plummer-equivalent force softening scale ($\epsilon$) is 360 pc~$h^{-1}$. The star-formation histories (SFHs) and stellar masses associated with each halo and subhalo were predicted using \textsc{UniverseMachine} DR1 \citep[UM-DR1;][]{UM2019, UM2021}, an empirical galaxy--halo connection model.

The properties of each of the 49 SymphonyGroup host halos, along with their associated subhalos, are saved at 73 snapshots within the redshift range of interest, $ 1 < z < 3.5$. We include a given host in our analysis at every snapshot where it satisfies the following conditions: 1) its \textsc{UniverseMachine} stellar mass is $\Mstar \geq 10^{10}~\Msun$, and 2) its specific star-formation rate is sSFR $\leq 10^{-10.5}~\mathrm{yr}^{-1}$. This is consistent with the criteria used to select the massive, quiescent hosts in the observational data, resulting in a total of 1344 hosts in our simulation sample. 

\begin{figure}
    \centering
    \includegraphics[width=0.47\textwidth]{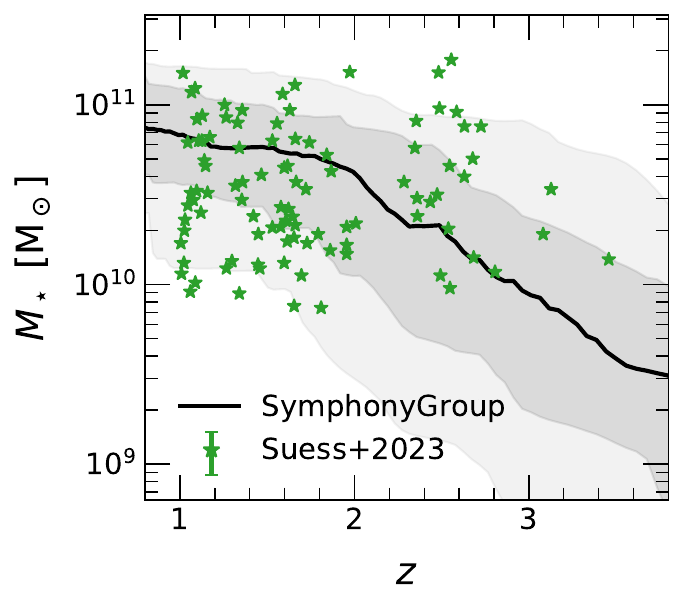}
    \caption{The distribution of stellar mass as a function of redshift of the SymphonyGroup and \citet{Suess2023} hosts. The median stellar mass history of SymphonyGroup is shown by the solid black line, and the dark (light) gray shaded region represents the 16th$-$84th (2.5th$-$97.5th) percentile range. The observed hosts from \cite{Suess2023} are overplotted with green stars; a characteristic error bar is shown in the legend.}
    \label{fig:host_distribution}
\end{figure}

Figure \ref{fig:host_distribution} plots stellar mass as a function of redshift for both the observed and simulated hosts. We note that while the majority of the observed hosts fall within the mass distribution represented by the SymphonyGroup hosts, there is a slight mismatch at the low-mass, low-redshift end ($z < 2$) and the high-mass, high-redshift end ($z > 2$). This stems from the fact that the SymphonyGroup halos were selected to exhibit a sharply peaked halo mass distribution at $z \lesssim 1$, which is not fully representative of the halo population associated with the observed host galaxies considered here. This underlying difference in halo mass distributions could consequently lead to the discrepancy in stellar mass between the observed and simulated hosts. We address this issue in Section \ref{sec:mass distribution} and Appendix \ref{app:mass distribution}.

The subhalos of each host are identified and characterized (assigned masses, positions, etc.) using the halo catalogs produced by two halo finders: \textsc{Rockstar} \citep{Rockstar2013} in the simulation snapshots prior to a subhalo crossing within the virial radius of its host (``pre-infall''), and \textsc{Symfind} \citep{Symfind2024} thereafter. Importantly, \textsc{Symfind} is a particle tracking-based halo finder that identifies the particles associated with a subhalo prior to infall and directly tracks these particles forward in time. This allows \textsc{Symfind} to follow subhalos down to orders-of-magnitude lower masses than those typically resolved by commonly used halo-finding tools, which typically rely on local density contrasts that can be quickly washed out within the host’s complex density field. The enhanced sensitivity of \textsc{Symfind} is crucial for studying simulated low-mass satellite galaxies. Stitching the \textsc{Symfind} and \textsc{Rockstar} catalogs together is necessary as \textsc{Symfind} depends on the existing \textsc{Rockstar} halo catalog to identify and track the particles of a subhalo prior to infall. \textsc{Symfind} then re-identifies and tracks the subhalo post-infall using only its own particles (see Section 3 of \citealt{Symfind2024} for details.) 

To mimic the 2D aperture used to identify satellites around hosts in the observations, we include a simulated subhalo if it falls within a 2D projected radius of 35 physical kpc. In other words, a simulated subhalo is considered an observed satellite if it falls within what is effectively a cylinder of radius 35 kpc and length $R_{\rm{vir,host}}$ along a given viewing direction. For each host, we construct satellite populations along the simulations' $x$- $y$-, and $z$-axes. All satellite counts reported in this paper are averages over these three viewing projections. The same mass-completeness limits are applied to the simulations and to the observations.

%%%%%%%%%%%%%%%%%%%%%%%%%%%%%%%%%%%%%%%%%%%%%%%%%%%%%%%%%%%%%%%%%%%%%%%%%

\section{Methods} \label{sec:methods}

\begin{figure*}
    \centering
    \includegraphics[width=0.98\textwidth]{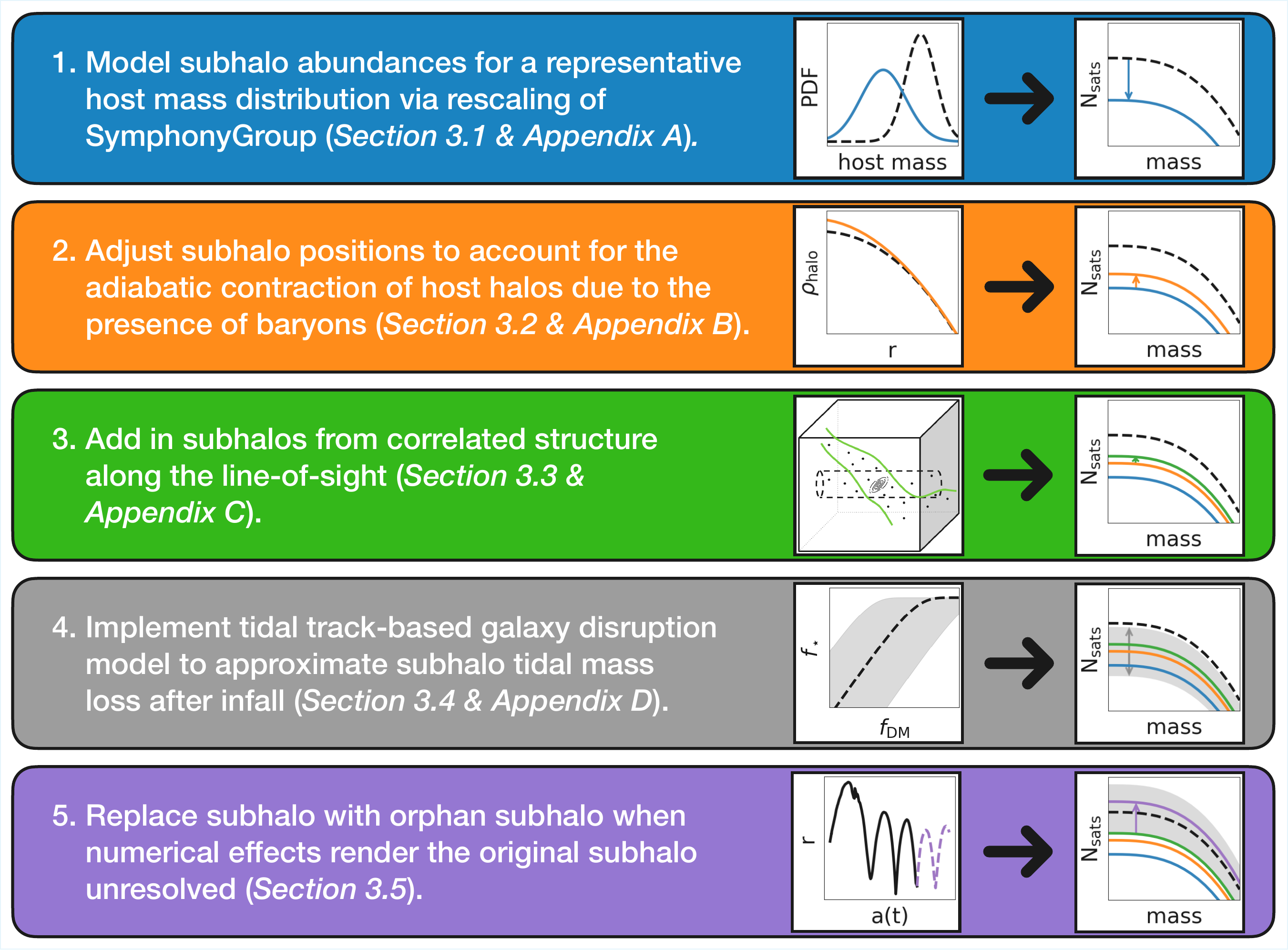}
    \caption{A schematic of the modeling steps implemented in this study to accurately compare simulated satellites to observations. These procedures are applied to every host halo at each relevant snapshot between $1 < z < 3.5$. Each panel corresponds to an individual modeling step and refers to a distinct subsection in Section \ref{sec:methods}. In each panel, the first plot illustrates the effect being modeled or the model component itself; the second plot shows its effect on the satellite stellar mass function.
    }
    \label{fig:methods}
\end{figure*}

In this section, we describe the modeling steps needed to accurately compare the simulated satellite population to that measured by observations. Figure \ref{fig:methods} presents a summary of these steps.

\subsection{Correcting the host mass distribution mismatch between observations and SymphonyGroup}
\label{sec:mass distribution}

To model the impact of the biased host mass distribution of SymphonyGroup, we assume that at a given redshift, the subhalo mass function at different host masses is ``self-similar.'' In this context, self-similarity means that the number of subhalos within some scaled projected radius $r/R_{\rm vir,host}$ and above some scaled mass $m/M_{\rm vir,host}$ is independent of host mass. The assumption of self-similarity is well-motivated by previous studies --- see, for example, Figure 3 in \citealt{BBK2017} and Figures 7 and 8 in \citealt{Nadler2023}. However, it is not exact: deviations from perfect self-similarity exist and represent a meaningful source of systematic uncertainty in our analysis, which we discuss further in Appendix \ref{app:mass distribution}.

Despite these caveats, assuming self-similarity allows us to scale the subhalo distributions of the SymphonyGroup hosts to approximate those of a representative host sample. This unbiased sample of hosts is selected from the ``Chinchilla'' Lb250-2048 cosmological box \citep{Lehmann2017} using the same mass and sSFR cuts mentioned in Section \ref{sec:simulations} ($\Mstar \geq 10^{10}~\Msun$ and sSFR $\leq 10^{-10.5}$ yr$^{-1}$).\footnote{We cannot simply measure the satellite mass functions because the resolution of the Chinchilla simulation box is too low to resolve subhalos down to the $\sim10^{7}~\Msun$ satellites probed by the observations.} We then translate this approximated subhalo population into satellite galaxies using the best-fit UM-DR1 satellite stellar–halo mass relation \citep[SHMR;][]{UM2019}, while preserving the original offsets of the SymphonyGroup subhalos from the median SHMR. If uncorrected, the distribution shift between the observed and simulated host populations can lead to a $\sim50\%$ difference in satellite abundances, but we estimate that the corrected shift leads to uncertainties less than $\sim10-20\%$. 

%In the context of mass functions, the assumption of self-similarity means that we assume that the average number of subhalos within some radius $r/R_{\rm vir,host}$ and above some mass $m/M_{\rm vir,host}$ is independent of host mass. Self-similarity is well-motivated by previous results; see, e.g., Figure 3 in \citealt{BBK2017} and Figures 7 and 8 in \citealt{Nadler2023}, although as can be seen in the above papers and as we discuss further in Appendix \ref{app:mass distribution}, this assumption is not completely correct, and that difference is a meaningful source of systematic uncertainty in our results. Assuming self-similarity allows us to scale the subhalo distributions of the SymphonyGroup hosts to approximate those of a representative host sample. We then translate these subhalos into satellite populations using the best-fit UM-DR1 satellite stellar–halo mass relation \citep[SHMR;][]{UM2019}, preserving the original offsets of the SymphonyGroup satellites from the median SHMR. The unbiased sample of hosts is selected from the ``Chinchilla'' Lb250-2048 cosmological box \citep{Lehmann2017} using the same mass and sSFR cuts mentioned in Section \ref{sec:simulations} ($\Mstar \geq 10^{10}~\Msun$ and sSFR $\leq 10^{-10.5}$ yr$^{-1}$).\footnote{We cannot simply measure the satellite mass functions because the resolution of the Chinchilla simulation box is too low to resolve subhalos down to the $\sim10^{7}~\Msun$ satellites probed by the observations.}

We additionally scale the overall redshift distribution of the Chinchilla hosts to match that of the observed hosts, and do the same for their stellar mass distributions in each of the redshift bins used in our analysis (see Table \ref{tab:observed hosts}). This allows us to assign weights to the simulated host--satellite systems, ensuring that any discrepancies between the observed and simulated satellite abundances are not due to distribution shifts between the two host populations' redshifts or stellar masses. A detailed description of all the methodology presented here can be found in Appendix \ref{app:mass distribution}.

Ideally, we would like to have a representative collection of hosts to begin with, rather than relying on post-hoc modeling to estimate their subhalo distributions. This can be achieved by running additional simulations that span the full mass range necessary to match observations ($\sim 10^{11.5-14}~\Msun$; see Figure \ref{fig:mass_distribution_shift} in Appendix \ref{app:mass distribution}). Such simulations are planned for future work.

\subsection{Accounting for the effect of baryons on host halo central densities}
\label{sec:adiabatic contraction}

Although on average dark matter exceeds baryonic matter by a factor of $\Omega_{\rm{DM}}/\Omega_{b} \approx 5.4$ \citep{Planck2020}, stars contribute significantly to the gravitational field in the central regions of galaxies \citep{Gnedin2004}. As baryons cool and condense at the centers of halos to form stars, they increase the central density and pull the dark matter inwards. This effect serves to elevate the subhalo counts within a given radial distance from the host. 

We apply adiabatic contraction \citep{Blumenthal1986} to the SymphonyGroup hosts and adjust the positions of subhalos within each host to account for this expected enhancement. Specifically, we numerically solve for the new radii of subhalos given the increased central potential, using the equation
\begin{equation}
    (1 + f_b)~M_{\mathrm{DM}}(r_i)~r_i = \big[ M_{\mathrm{DM}}(r_i) + M_b (r_f) \big]r_f,
\end{equation}
for $M_{\mathrm{DM}}(r_f)$, the final contracted host dark matter distribution. Here, $M_{\mathrm{DM}}(r_i)$ is the initial host halo profile, $M_b(r_f)$ is the final baryon profile, and particles at initial radius $r_i$ are contracted to $r_f$. The adiabatic contraction model assumes spherical symmetry, circular orbits, conservation of angular momentum and homologous contraction (or ``no shell-crossing''). We measure the initial host halo profiles directly from the simulations and model the final baryonic distribution using a deprojected S\'ersic profile. Assuming the locations of subhalos contract in the same way as their underlying halos, we can then adjust their positions such that a subhalo initially separated from its host by $r_i$ now sits at $r_f$. 

We present a comprehensive description of our adiabatic contraction implementation and quantify its effect on the satellite stellar mass function in Appendix \ref{app:adiabatic contraction}. We find that the inclusion of adiabatic contraction increases satellite counts by $\sim4\%$ at $z\sim 1$; this effect decreases to $\lesssim 2\%$ at $z \gtrsim 2$. 

We note that the baryonic mass content of host halos also exerts tidal forces on subhalos and can accelerate subhalo tidal mass loss \citep{BrooksZolotov2014, Sawala2017, GarrisonKimmel2017, Barry2023, EDEN2024}. This effect can be important, although it is expected to be less important for the satellites of high redshift massive systems than it is for low redshift disk galaxies like the Milky Way. While this effect is not included in the present analysis, it will be explored in future work.

\subsection{Adding in the contribution of interlopers from correlated structure}
\label{sec:interloper modeling}

The hosts relevant to our study are massive, quiescent galaxies surrounded by numerous satellites, indicating that they are found in relatively high-density environments \citep[e.g.,][]{Peng2010, Tal2013}. Thus, compared to a random patch of sky, these objects are more likely to be in the vicinity of a larger-scale structure such as a filament. In this case, galaxies associated with the correlated structure might appear to be part of the host system when projected on the sky. Such galaxies are challenging to distinguish from true satellites in the observations, as their redshifts are often indistinguishable within measurement uncertainties. Therefore, we model this contamination using simulations and add these ``correlated interlopers'' to our simulated subhalo counts to enable a fair comparison with the observed satellites. 

We estimate the contribution of correlated interlopers by comparing galaxy abundances along host-centered sightlines to those along random sightlines, using the Chinchilla Lb250-2048 cosmological simulation \citep{Lehmann2017}. At each snapshot between $1 < z < 3.5$ in the simulation, we select 5000 host galaxies with stellar masses above $10^{10}\Msun$ and define a cylindrical sightline of radius 35 kpc extending along the $\hat{x}$, $\hat{y}$, and $\hat{z}$ directions from each host. We count the number of galaxies above a series of stellar mass thresholds as a function of line-of-sight distance from the host, starting at the virial radius and extending to 125 Mpc $h^{-1}$. This yields the number of non-satellite galaxies projected into our line of sight. Then, to isolate the contribution from correlated structure, we define 5000 random sightlines throughout the simulation volume and repeat the same galaxy counting procedure. The difference between the host-centered and random sightline counts (averaged over viewing direction) gives a direct estimate of the number of correlated interlopers as a function of stellar mass threshold and redshift.

We find that the relationship between the number of correlated interlopers $\Nint$, scale factor $a(t)$, and minimum stellar mass threshold $m \equiv \log_{10}M_\star$ is well fit by the following equation:
\begin{equation}
    \log_{10} \Nint(a, m) = \alpha(m) \log_{10}\big(a(t)\big) + \beta(m),
\end{equation}
where $\alpha(m)$ and $\beta(m)$ are given by
\begin{equation}
    \alpha(m) = 9.392 - 3.259 m + 0.228 m^2,~\rm{and}
\end{equation}
\begin{equation}
    \beta(m) = -0.205 - 0.292 m.
\end{equation}
The details of this modeling can be found in Appendix \ref{app:interloper modeling}. Correlated interlopers are found to contribute between $\sim 0.01\%$ to the predicted satellite SMFs at $z \sim 1$ and $\sim0.1\%$ at $z \sim 3$.

\subsection{Modeling the tidal evolution of subhalos}
\label{sec:subhalo mass loss}

As a subhalo falls into the gravitational potential of its host, it experiences destructive tidal forces that gradually strip mass away from its outskirts in a process known as tidal stripping \citep{Moore1996, Tormen1998, Behroozi2014, vdBosch2018}. The evolution of a subhalo's physical properties (e.g., radius, peak circular velocity) as it undergoes tidal stripping seem to be independent of the details of its orbit; instead, the evolution of these characteristic parameters depend solely on the total amount of mass lost, following well-defined ``tidal tracks'' \citep{Penarrubia2008, Smith2016, ErraniNavarro2021}. This picture stands in contrast to the most common prescription for modeling galaxy mass loss in simulations, which we refer to here as the ``disruption model.''

In the disruption model, a satellite loses its stars in a step-function manner. It stays perfectly intact until some cataclysmic moment of total disruption, after which it has no stars. Each model has its own scheme for predicting when this obliteration occurs (for example, in \textsc{UniverseMachine}, it happens after the central rotation curve of the subhalo has dropped below some fitted threshold; \citealt{UM2019}). However, for most satellites, stellar mass loss is more gradual and is significantly intertwined with the detailed properties of the galaxy. While the SymphonyGroup zoom-ins lack the resolution and hydrodynamics needed to explicitly model the stars, we can model the mass loss of the subhalos, which are orders of magnitude more massive than the stellar component. Because of this, we adopt a tidal track-based mass loss model to predict the evolution of stellar masses after infall, wherein a galaxy’s remaining stellar mass is determined by its remaining subhalo mass. 

The most direct way to study the relationship between stellar and dark matter tidal mass loss is through hydrodynamic simulations, which explicitly model both the baryonic and dark matter components of subhalos. This approach was taken by \citet{Smith2016}, who found that, indeed, the bound dark matter fraction of a subhalo alone serves as a good predictor of its bound stellar fraction, with galaxy mass loss accelerating exponentially once the subhalo falls below a characteristic mass scale. We adopt the tidal track parametrization of \citet{Smith2016} to describe galaxy tidal stripping in this work. The analytic form of this tidal track is given by
\begin{equation}
    \fstr = 1 - \exp\left[\ln\Big(\frac{1}{2}\Big) \frac{\fdm}{f_{1/2}}\right],
    \label{eq:tidal_track}
\end{equation}
where $f_{\rm{DM}} \equiv m_{\rm{DM}}/m_{\rm{DM,infall}}$ is the bound fraction of a subhalo's dark matter mass, $f_\star \equiv m_\star/m_{\rm{\star,infall}}$ is its bound stellar mass fraction, and $f_{1/2}$ is the value of $f_{\rm{DM}}$ at which $f_\star = 1/2$. Stellar masses for subhalos can then be estimated by calculating $m_\star = f_\star \cdot m_{\rm{\star,infall}}$ at every post-infall snapshot. Here, $f_{1/2}$ is the key parameter that sets how quickly stellar stripping becomes significant relative to dark matter stripping. A value of $f_{1/2} \sim 1$ means that a galaxy loses half its stellar mass immediately upon infall, while a value of $f_{1/2} \ll 1$ indicates that most of a subhalo's dark matter must be lost before its stars are appreciably affected. The tidal tracks of the simulated galaxies studied in \citet{Smith2016} were best fit by $f_{1/2} = 0.0488$. We adopt this value of $f_{1/2}$ in our fiducial model\footnote{$f_{1/2}$ is related to the $a_{\rm{strip}}$ parameter of the \citet{Smith2016} tidal track parameterization via $f_{1/2} = -\ln(1/2)/a_{\rm{strip}}$.}, although we note that different theoretical models predict very different values of this quantity (see below).

Implicit in this choice of constant $f_{1/2}$ is the assumption that the stellar-to-halo size ratio is fixed across the galaxy population, and thus, all galaxies evolve along the same tidal track. To first order, this appears reasonable (see, e.g., Figure 3 of \citealt{Smith2016}). However, there are a couple of caveats. First, there is known scatter in this ratio, and the compactness of a galaxy relative to its halo is a key factor that can impact the efficiency of stellar tidal stripping. The more deeply embedded the stars are within a halo, the more resistant they are to being stripped \citep{Penarrubia2008, Smith2016}. In other words, as the stellar-to-halo size ratio decreases, so does $f_{1/2}$. Additionally, galaxy sizes \citep[e.g.,][]{vanderWel2014, Shibuya2015} and baryon content \citep[e.g.,][]{Genzel2017, Lang2017, Walter2020} are known to evolve with redshift at fixed stellar mass, which could lead to redshift-dependent variation in $f_{1/2}$. 

Furthermore, the \citet{Smith2016} tidal track functional form is empirically determined using hydrodynamic simulations, which are subject to numerical artifacts that preferentially heat the stellar distributions of simulated galaxies and cause spurious size growth \citep{Ludlow2019, Ludlow2020, Ludlow2021, Ludlow2023}. Because larger galaxies lose stars to tides more rapidly (at a fixed halo size), the mischaracterization of galaxy sizes in hydrodynamic simulations can result in the mischaracterization of stellar stripping. Idealized models that do not suffer from the same numerics tend to predict a lower $f_{1/2}$ value than current hydrodynamic simulations \citep[e.g.,][]{Errani2022}. 

In this analysis, we explore these uncertainties in satellite size and disruption physics by allowing $f_{1/2}$ to vary between $10^{-3}$ and 0.5. This range of $f_{1/2}$ values is likely excessively wide --- we do not expect a subhalo to have already lost half its stars by the time half of its dark matter has been stripped (the $f_{1/2} = 0.5$ case). We do, however, expect that this broad range brackets the true scope of $f_{1/2}$ and encompasses the full range of disruption physics consistent with hydrodynamic simulations ($f_{1/2} \approx 0.0290 - 0.0806$; \citealt{Smith2016}) and idealized models ($f_{1/2} \approx 0.001 - 0.003$; see Appendix \ref{app:galaxy profiles}). The idealized model predictions for $f_{1/2}$ come from the tidal disruption model of \citet{Errani2022}, which we compare to our fiducial model in Appendix \ref{app:galaxy profiles}. While the shape of our fiducial tidal track closely resembles that of the \citet{Errani2022} model for a galaxy with an $n = 2$ S\'ersic profile, the idealized tidal tracks are systematically shifted towards later disruption times. This is reflected in their significantly lower $f_{1/2}$ values, over an order of magnitude below those measured from the hydrodynamic simulations that inform our fiducial model. 

We explore the impact of the assumed stellar profile on simulated subhalo abundances and the resulting constraints on $f_{1/2}$ in Appendix \ref{app:galaxy profiles}. We find that the choice of stellar profile shape can shift the inferred value of $f_{1/2}$ by $\sim60\%$, underscoring the significant role that galaxy structure plays in tidal evolution.

\subsection{Replacing numerically unreliable subhalos with ``orphan'' subhalos}
\label{sec:numerics + orphan model}

To compare simulation predictions to observations of satellite galaxies, one must be able to identify the subhalos that host those galaxies in simulations. This task is complicated by the fact that not all physically surviving subhalos are faithfully tracked in cosmological simulations. Artificial subhalo disruption can arise from numerical or methodological artifacts, resulting in a systematic underestimation of the surviving subhalo population. For this reason, it is common for simulations to implement an ``orphan model'', which approximates the properties of subhalos that are prematurely disrupted. In this section, we summarize the key sources of artificial disruption in numerical simulations (Section \ref{sec:numerical disruption}) and describe the orphan model adopted in this work to mitigate their impact (Section \ref{sec:orphan model}).

\subsubsection{What causes artificial disruption?}
\label{sec:numerical disruption}

The disappearance of subhalos from simulation outputs (or ``halo catalogs'') is often not due to true disruption resulting from physical processes that unbind all the mass within a subhalo, but instead to numerical or methodological limitations. First of all, halos and subhalos are not predefined structures in an $N$-body simulation; rather, they emerge from the gravitational interactions between the simulated dark matter particles. Thus, post-processing is necessary to identify where the particles have collapsed into bound structures at each snapshot of the simulation. This is achieved using ``halo finders'', which detect halos and subhalos by identifying local overdensities with various methods \citep[e.g.,][]{Knebe2011}. However, halo finding is challenging, as subhalos can easily get lost within the complex background density of their host halos \citep{Knebe2011}. This challenge is compounded by the need to consistently identify the same (sub)halos across multiple simulation snapshots, including during periods of strong tidal stripping. As a result, halo finders frequently mischaracterize subhalos --- assigning incorrect masses, misidentifying their positions in subsequent snapshots, or both \citep{Symfind2024}.

To make matters worse, numerical effects can lead to artificial premature disruption. The fundamental problem with simulating subhalos is that the continuous dark matter field needs to be discretized into a finite number of particles to make the calculation tractable. This discretization necessitates the use of force softening, which assigns an unphysical size and density profile to each particle to prevent spurious gravitational interactions; causes subhalo mass loss to become a discrete rather than continuous process; and allows for unphysical two-body interactions between simulation particles. 

All three effects lead to significant numerical challenges. Force softening smooths the gravitational potential of subhalos on small scales, suppressing their central densities and making them more vulnerable to stripping \citep[e.g,][]{vdBoschOgiya2018, Ludlow2019a}. Discreteness noise induces stochastic mass loss events, and the response of a subhalo to losing too few particles is asymmetric with its response to losing too many: the former has little impact on the subhalo's structure and future evolution, while the latter causes future mass loss to be more rapid and can lead to runaway disruption \citep{vdBoschOgiya2018}. Two-body scattering also allows artificial energy exchange between particles, leading to numerical relaxation that gradually transforms the inner regions of a subhalo from a realistic dark matter profile into a non-physical, collision-dominated profile \citep[e.g.,][]{Power2003,Ludlow2019a,Symfind2024}. Furthermore, the size of this numerically relaxed region grows with time, and once tides remove enough of the subhalo's mass, only the collisional core remains --- at which point its mass loss rate becomes unreliable or unconverged. 

These modeling limitations systematically make subhalos appear more fragile in cosmological simulations than they truly are, resulting in artificial disruption. 

\subsubsection{The orphan model}
\label{sec:orphan model}

To account for subhalos lost due to numerical disruption or limitations of the halo finder, it is common to track subhalos until they disappear from the halo catalog, and then apply a post-processing orphan model to estimate their orbits and mass evolution thereafter. While \textsc{Symfind} is able to reliably track long-lived, low-mass subhalos, the properties associated with those subhalos are not guaranteed to be free of numerical effects. According to the reliability tests in \citet{Symfind2024}, the numerical effects discussed in Section \ref{sec:numerical disruption} begin to bias a subhalo's mass loss rate when the subhalo's particle count falls below $n < n_{\rm{lim}}$, where $n_{\rm{lim}}$ is well fit by the relation
\begin{equation}
    n_{\rm{lim}}(n_{\rm{peak}}) = 10^{-0.01853x^2 + 0.3861x + 1.6597} \cdot n_{\rm{peak}},
    \label{eq:nlim}
\end{equation}
where $n_{\rm{peak}}$ is the subhalo peak particle count and $x \equiv \log_{10}(8n_{\rm{peak}})$. Thus, once the particle count of a low-mass subhalo falls below $n_{\rm{lim}}$, we replace it with an orphan subhalo, irrespective of \textsc{Symfind}'s ability to find it. This marks a significant improvement over conventional orphan models, which initiate orphans only after a subhalo becomes undetectable, without regard for its numerical reliability prior to disappearing. %Moreover, orphan subhalos in our model are not manually removed; they simply lose mass until they naturally fall below the mass-completeness limits. By bypassing the need to choose when to disrupt an orphan subhalo, we eliminate a key source of systematic uncertainty.

The orbit of an orphan subhalo is estimated using the position and velocity of the subhalo's particle that was most bound at infall \citep[e.g.,][]{White1987, Wang2006, Pujol2017}. Highly bound particles are stripped from subhalos later than lightly bound particles, so the most-bound particle is a simple method for finding a tracer that follows the orbit of the subhalo. Orphan models often identify the most-bound particle in the snapshot before a subhalo disrupts; however, as shown in \cite{Symfind2024}, this exposes such models to significant numerical errors. Using the particle data allows us to avoid making simplifying approximations, for example, that the host halo follows a smooth profile, which would be required for analytically integrating a test particle's orbit forward. We verified that the subhalo orbits determined by their most-bound particles at infall agree to within $\sim1.5\%$ of their true \textsc{Symfind}-determined orbits. Thus, implementing this orphan model allows us to estimate the true physical properties of subhalos even after they are subject to numerical effects.

To evaluate the mass of an orphan, we follow the subhalo mass loss prescription used by the \textsc{UniverseMachine} orphan model, based on the work of \citet{Jiang2016} (see Appendix B of \citealt{UM2019}). Here, subhalos only lose mass after each pericenter passage:
\begin{equation}
    \dot{m}_{\rm{infalling}} = 0
\end{equation}
\begin{equation}
    \dot{m}_{\rm{outgoing}} = -1.18 \frac{m_{\rm{sub}}}{t_{\rm{dyn}}} \Big( \frac{m_{\rm{sub}}}{M_{\rm{host}}}\Big)^{0.07},
\end{equation}
where $m_{\rm{sub}}$ is the subhalo virial mass, $M_{\rm{host}}$ is the host virial mass, and $t_{\rm{dyn}} \equiv \big( \frac{4}{3} \pi G \rho_{\rm{vir}} \big)^{-1/2}$ (i.e. the crossing time). Whether a subhalo is infalling or outgoing is determined by its radial velocity. At each snapshot, we estimate the mass loss rate $\dot{m}$ and assume it remains constant over the interval until the next snapshot, allowing us to compute the total mass lost.

We used this model to estimate the post-infall masses of a sample of resolved subhalos and find that, on average, it is able to predict their true mass histories to within $\sim$0.2--0.3 dex and with minimal bias. That said, this model for mass loss is necessarily incomplete. This is an orbit-averaged model and does not account for the fact that low-radius subhalos in strong tidal fields will always lose mass more rapidly than their high-radius counterparts. The $\sim$0.2--0.3 scatter that we see is almost certainly physical scatter caused by marginalizing over all tidal field strengths and will be reducible with a more sophisticated model. We experimented with several alternative mass-loss estimation schemes, such as fitting a curve to the converged mass loss and extrapolating, but were unable to reduce the scatter below that of the \citet{Jiang2016} model.
Within the projected 35 kpc radius used to identify satellites, we find that the orphan model increases satellite counts by a factor of $\sim1.8$ at lower resolutions ($n_{\mathrm{peak}} \lesssim 10^4$), and has no effect at high resolutions ($n_{\mathrm{peak}} \gtrsim 10^5$).

We note that our limits on when a stellar mass-selected population is free from numerical effects/orphan subhalos are more conservative than those presented for the \citet{Smith2016} disruption model in Figure 13 of \citet{Symfind2024}. At first glance, this may seem contradictory --- we use the \citet{Symfind2024} numerical reliability model to determine when to turn subhalos into orphans, and our fiducial galaxy disruption model comes from \citet{Smith2016}. However, this discrepancy arises because Figure 13 of \citet{Symfind2024} implicitly assumes step function-like disruption once subhalos pass $f_{1/2}$, while our model allows very heavily disrupted subhalos to continue contributing to the satellite stellar mass function. At the small radii probed in this study, a substantial fraction of our satellite galaxies have experienced significant mass loss and can have $m/m_{\rm{peak}}$ values more than an order of magnitude below $f_{1/2}$, which makes our resolution requirements significantly stricter. This difference is a strong illustration of the interplay between galaxy disruption models and numerical completeness limits.

%%%%%%%%%%%%%%%%%%%%%%%%%%%%%%%%%%%%%%%%%%%%%%%%%%%%%%%%%%%%%%%%%%%%%%%%%

\begin{figure*}[ht]
    \centering
    \includegraphics[width=0.98\textwidth]{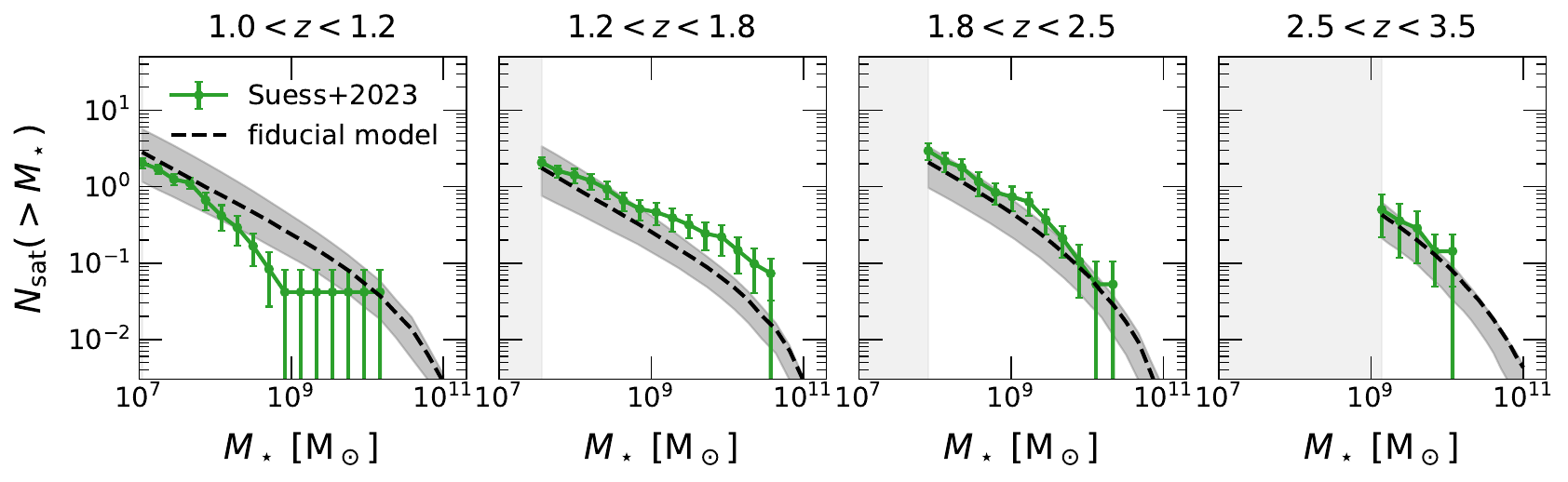}
    \caption{Comparison between theoretical and observed cumulative satellite mass functions in four bins of redshift. The solid green curves represent the satellite mass functions measured from \citet{Suess2023} and their statistical uncertainties, estimated using $10^5$ bootstrap resamples. The dashed black curves depict the theoretical mass functions derived from our fiducial model, which assumes $f_{1/2} = 0.0488$. The dark gray shaded regions represent the range of satellite mass functions allowed by $f_{1/2}$ values between 0.001 and 0.5. The light gray vertical bands indicate the mass range over which our observations are incomplete. The fiducial model slightly over-predicts the normalization of the satellite mass function at $1.0 < z < 1.2$ and under-predicts the normalizations at $1.2 < z < 1.8$ and $1.8 < z < 2.5$, and is fully consistent with the fiducial model at $2.5 < z < 3.5$.}
    \label{fig:mass_function_f0}
\end{figure*}

\section{Results} \label{sec:results}

\subsection{Satellite stellar mass functions}

We compare the cumulative mass functions of the JWST-observed satellites from \citet{Suess2023} against the simulated satellites in Figure \ref{fig:mass_function_f0}. The comparison is shown in four bins of redshift between $1.0 < z < 3.5$, chosen to match the bin spacing of the mass-completeness limits of the observations (see Section \ref{sec:observations}). In each bin, we compare the satellite mass functions measured from \citet{Suess2023}, along with their statistical uncertainties estimated via bootstrapping ($N_{\rm{bootstraps}} = 10^5$) over the observed hosts, to the fiducial theoretical predictions. The latter assumes $f_{1/2} = 0.0488$, following \citet{Smith2016}; the range of the model shown in the figure represents the range of satellite mass functions allowed by $f_{1/2}$ values between 0.001 and 0.5. The light gray vertical shaded bands indicate the regions below the mass-completeness limits of the observations. 

We first consider the satellite stellar mass functions predicted by our model. The theoretical lower bounds on the SMFs are determined by setting $f_{1/2} = 0.5$ in our tidal evolution model. This limit was chosen as an extreme case, describing highly fragile galaxies that disrupt near-instantaneously. To our knowledge, no existing galaxy models explicitly predict such fragility. Moreover, any value of $f_{1/2} > 0.5$ would imply satellites lose half their stars \textit{before} losing half their dark matter, which is physically implausible, implying 0.5 is a safe upper limit. Conversely, the SMF upper bounds are set by assuming $f_{1/2} = 0.001$. For this value, satellite galaxies are extremely durable --- they can lose over 99\% of their dark matter before half their stellar mass is lost. Within this framework, the range of allowable satellite SMFs spans a factor of $\sim$3--3.7, over half an order of magnitude. Even within the more physically motivated range of $0.001 \leq f_{1/2} \leq 0.1$, the SMF still varies by a factor of $\sim$1.5--2, highlighting the extent of the systematic uncertainty imparted on the predicted satellite SMF from the uncertainty in modeling galaxy disruption physics. 

We find that the observed satellite stellar mass functions between $1.0 < z < 3.5$ are generally consistent with the fiducial model predictions within a factor of $\sim$ 2--3. Given the observational uncertainties, this translates to a 1--2$\sigma$ agreement between the model and observations at all masses and redshifts. In the highest-redshift bin ($2.5 < z < 3.5$), the overall amplitude of the observed and fiducial model SMFs are in full agreement. In contrast, at intermediate redshifts ($1.2 < z < 1.8$ and $1.8 < z < 2.5$), the model tends to slightly under-predict the mass function normalization. In particular, in the $1.8 < z < 2.5$ bin, the observed satellite SMF sits near the upper bound of the range permitted by our galaxy tidal evolution model, requiring galaxies to be extremely resistant to tidal stripping to reproduce the observed satellite population. Lastly, in the lowest-redshift bin ($1.0 < z < 1.2$), the normalization of the observed stellar mass function is lower than that predicted by the fiducial model, suggesting that satellites may be less durable than assumed.

Additionally, the slope of the observed $1.0 < z < 1.2$ satellite SMF appears to diverge from that of the model prediction between $\sim10^{8-10}~\Msun$. However, in cumulative mass functions like these, statistical fluctuations can masquerade as slope changes: in cumulative functions, fluctuations impact the amplitude of multiple consecutive bins. These statistical fluctuations may be the culprit, given that both the low- and high-mass ends of the observed SMF \textit{do} match the model prediction, and only 21 systems are contributing to the measurements in this redshift bin. The same argument can be made for the slope mismatch in the $1.2 < z < 1.8$ redshift bin.

Using the satellite SMFs as a starting point, we can ask two questions: \textit{Does the model reproduce the same redshift trend as the observations?} and \textit{Does the model reproduce the same overall satellite abundance as the observations?} We discuss these two questions in the subsequent sections.

\subsection{Redshift evolution of satellite counts}

As previously mentioned, our fiducial model assumes that $f_{1/2}$ is a constant parameter, i.e., it does not vary with redshift. However, this need not be the case. In fact, a redshift-dependent $f_{1/2}$ is plausible given that both the sizes and baryon content of galaxies are known to evolve with redshift \citep[e.g.,][]{vanderWel2014, Genzel2017}. Moreover, as shown in Figure \ref{fig:mass_function_f0}, the fiducial model sightly over-predicts the normalization of the satellite mass function at $1.0 < z < 1.2$ and under-predicts it in the $1.8 < z < 2.5$ bin, providing additional indications that $f_{1/2}$ may evolve with redshift.

\begin{figure}[ht]
    \centering
    \includegraphics[width=0.47\textwidth]{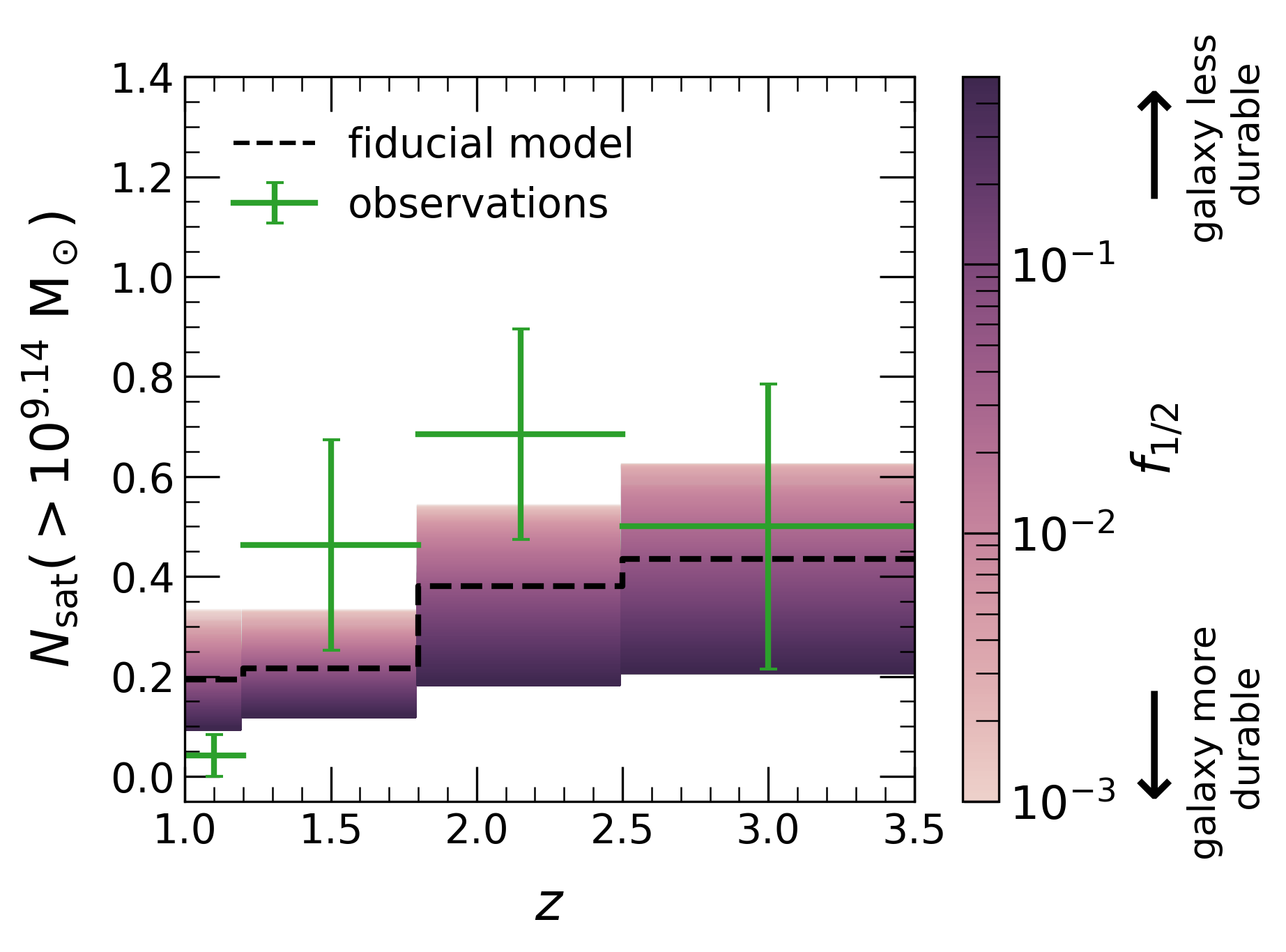}
    \caption{Average number of satellites per host above $10^{9.14}~\Msun$ (the mass-completeness threshold of the highest-redshift bin) as a function of redshift bin. The satellite abundances measured from the \citet{Suess2023} observations are shown in green. The error bars mark the 16th--84th percentile range obtained by bootstrapping over the hosts ($N_{\rm{bootstraps}} = 10^5$). The color-gradient bars represent the range of satellite counts predicted by our model framework, where the darkness of the color tracks the value of $f_{1/2}$ (i.e., galaxy durability). The black dashed line marks the total satellite abundance predicted by the fiducial model, which assumes a constant value of $f_{1/2} = 0.0488$ across all redshifts. To reproduce the redshift evolution in satellite abundance seen in the observations, $f_{1/2}$ may need to be a redshift-dependent parameter. However, the observational uncertainties are large, and the data remain consistent with the model within 1--2$\sigma$.}
    \label{fig:nsat_vs_z}
\end{figure}

We further investigate the idea of redshift evolution in Figure \ref{fig:nsat_vs_z}, which plots the average total abundance of observed and predicted satellites per host across the four redshift bins from Figure \ref{fig:mass_function_f0}. We impose a minimum satellite stellar mass $10^{9.14}~\Msun$, which is the mass-completeness threshold of the highest-redshift ($2.5 < z < 3.5$) bin. The satellite counts measured from the \citet{Suess2023} data are shown in green, with error bars indicating the bootstrapped 16th--84th percentile range (with $N_{\rm{bootstraps}} = 10^5$). The color-gradient bars represent the range of satellite counts predicted by our model for $10^{-3} \leq f_{1/2} \leq 0.5$. The black dashed line marks the satellite abundances predicted by our fiducial model.

The observations suggest that the average satellite abundance around massive, quiescent host galaxies tends to increase with redshift (with large uncertainties at $z > 2.5$ due to a small number of hosts). The fiducial model predicts a qualitatively similar redshift evolution. The degree of this evolution, however, appears somewhat stronger in the observations than in our fiducial, constant-$f_{1/2}$ model. This may imply that $f_{1/2}$ is not a constant parameter, but rather decreases with increasing redshift. At $1.0 < z < 1.2$, the observations prefer a value of $f_{1/2}$ larger than the fiducial model ($f_{1/2} \gtrsim 0.5$), while at $1.2 < z < 2.5$, they prefer much lower values ($f_{1/2} \lesssim 0.01$). Additionally, at intermediate redshifts, the observations potentially indicate higher satellite abundances than even an evolving $f_{1/2}$ model allows. 

These discrepancies between the observations and the fiducial model, however, are likely exaggerated by statistical fluctuations in the observed satellite SMFs around the $\Mstar\sim10^9~\Msun$ scale (see Figure \ref{fig:mass_function_f0}). We also do not account for sample variance in the observed systems, which is likely non-negligible given the relatively small number of hosts ($\sim20$) in each redshift bin. Even so, the observational uncertainties are large, and the model predictions remain consistent with the observations at the $1-2\sigma$ level. Future observational studies that include a larger number of hosts and a detailed characterization of systematic biases are required to assess the significance of this disagreement and shed light on the redshift evolution of $f_{1/2}$.

\subsection{The overall satellite abundance}

\begin{figure}[ht]
    \centering
    \includegraphics[width=0.47\textwidth]{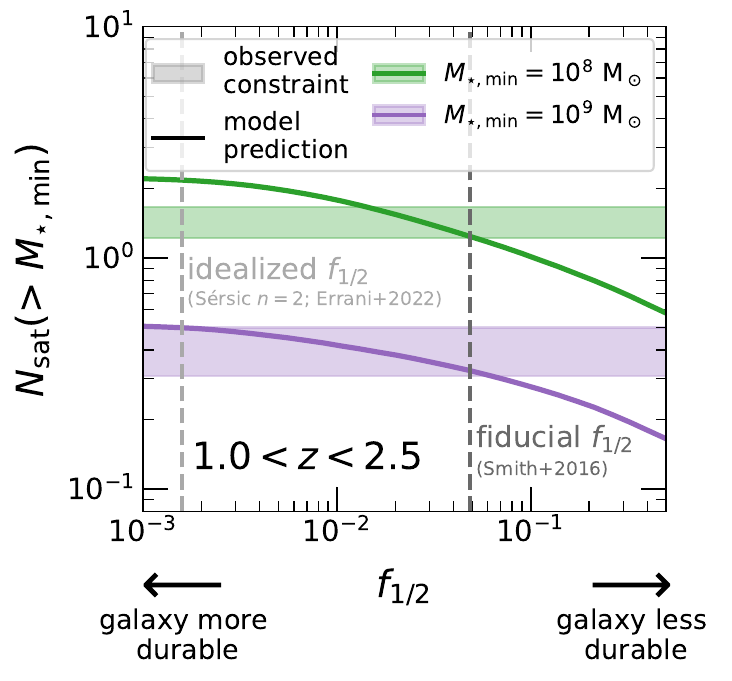}
    \caption{Impact of assumed galaxy mass loss model on the total simulated satellite abundances with stellar masses above two different mass thresholds between $1.0 < z < 2.5$. Note that we forgo the the highest-redshift bin ($2.5 < z < 3.5$) from the previous figures to maintain a lower mass-completeness limit. The solid green (purple) curve shows the average number of simulated satellites per host above $10^8~\Msun$ ($10^9~\Msun$) as a function of $f_{1/2}$. The green and purple shaded bands represent the corresponding satellite counts measured from the \citet{Suess2023} observations. The vertical dashed dark gray line marks our fiducial value of $f_{1/2} = 0.0488$. The light gray dashed line marks the value of $f_{1/2} = 0.0016$ predicted by the idealized \citet{Errani2022} model for a galaxy following a S\'ersic profile with $n = 2$. The observational bounds for both limiting stellar masses are consistent the fiducial value of $f_{1/2}$. The observed $N_{\mathrm{sat}} (\Mstar > 10^8~\Msun)$ appears to disfavor the idealized $f_{1/2}$ value, while the observed $N_{\mathrm{sat}} (\Mstar > 10^9~\Msun)$ remains compatible with it.}
    \label{fig:nsat_vs_f0}
\end{figure}

The average number of simulated satellites per host that fall above $10^8~\Msun$ and $10^9~\Msun$ as a function of $f_{1/2}$ in the redshift range $1.0 < z < 2.5$ is shown in Figure \ref{fig:nsat_vs_f0}. Note that here we are disregarding the highest-redshift bin in Figures \ref{fig:mass_function_f0} and \ref{fig:nsat_vs_z}, where the mass-completeness threshold becomes significantly higher. The shaded colored bands represent the bootstrapped 16--84th percentile range in the corresponding satellite counts measured from the \citet{Suess2023} observations. The results are compared to our fiducial value of $f_{1/2} = 0.0488$ and to the value of $f_{1/2} = 0.0016$ predicted by the idealized \citet{Errani2022} model for a galaxy following an $n = 2$ S\'ersic profile.

Over the $1.0 < z < 2.5$ redshift range, the number of satellites per host with $\Mstar > 10^8~\Msun$ and $\Mstar > 10^9~\Msun$ found in the \citet{Suess2023} observations --- $N_{\mathrm{sat}} \sim 1.4$ and $\sim 0.4$, respectively --- agree with our fiducial model. The $N_{\mathrm{sat}} (\Mstar > 10^8~\Msun)$ data are also consistent with values of $f_{1/2}$ as low as $\sim0.02$, suggesting that satellites may be 2--3 times more durable than predicted by hydrodynamic simulations. 

On the other hand, the implications for the idealized model prediction of $f_{1/2} = 0.0016$ are less clear. While the $N_{\mathrm{sat}} (\Mstar > 10^8~\Msun)$ observations appear to rule out this value of $f_{1/2}$ at the $2\sigma$ level, the $N_{\mathrm{sat}} (\Mstar > 10^9~\Msun)$ data is consistent with it. This highlights that our ability to constrain $f_{1/2}$ remains highly sensitive to the statistical uncertainties in the observed satellite stellar mass function --- the conclusions we draw about the viable values of $f_{1/2}$ vary depending on the chosen stellar mass threshold due to fluctuations in the satellite stellar mass function. The assumed galaxy profile shape in the idealized model (e.g., the choice of S\'ersic index $n$) also affects the resulting constraint on $f_{1/2}$, although its impact is subdominant (refer to Appendix \ref{app:galaxy profiles}). As a result, we cannot currently distinguish between different galaxy tidal mass loss models or robustly determine the true range of $f_{1/2}$. Future observational work with larger host samples and better-characterized uncertainties will improve our ability to do so.

Presently, we are limited by large observational uncertainties. The observational errors presented in Figures \ref{fig:mass_function_f0}, \ref{fig:nsat_vs_z}, and \ref{fig:nsat_vs_f0} are \textit{statistical} uncertainties only. Increasing the size of the observational sample is key to lowering these statistical uncertainties and smoothing out random variations in the satellite SMFs. Systematic biases, such as those introduced by random chance alignments, are not presently accounted for the observational error budget. \citet{Suess2023} estimated that random interlopers may bias satellite counts high by $\lesssim 20\%$, though additional work is necessary to fully characterize their influence on the observational data. Additionally, there are uncertainties in the galaxy--halo connection model that may limit our current ability to robustly constrain $f_{1/2}$. We discuss systematics that can impact the apparent consistency (or lack thereof) between our observed and predicted satellite abundances further in Section \ref{sec:systematics}. 

Nevertheless, it is noteworthy that these first-of-their-kind observations of low-mass companions around massive, quiescent galaxies at $1 < z < 3.5$ agree with our fiducial model within a factor of 2--3. In fact, the observed satellite mass function at $z \sim 1$ is fully consistent with our model prediction. This showcases both the predictive power of our current cosmological simulations, as well as the constraining power of high-precision JWST measurements in this previously unexplored mass and redshift regime.

%%%%%%%%%%%%%%%%%%%%%%%%%%%%%%%%%%%%%%%%%%%%%%%%%%%%%%%%%%%%%%%%%%%%%%%%%

\section{Controlling Systematics in Satellite Population Models} \label{sec:systematics}

In this section, we quantify several dominant sources of theoretical systematic uncertainty or bias in our modeling (Section \ref{sec:theory systematics}). We also assess additional uncertainties arising from the galaxy--halo connection (Section \ref{sec:GHC}) and from the observational data (Section \ref{sec:obs systematics}). At present, we do not directly model additional uncertainty arising from the central galaxies' impact on satellite disruption. Throughout, we discuss how these sources of uncertainty are controlled in our current analysis or how they can be addressed in future work.

\subsection{Theory systematics}
\label{sec:theory systematics}

As discussed in Section \ref{sec:methods}, we have made significant progress in understanding the main modeling components that contribute to the systematic uncertainty in our theoretical predictions about the satellite population of massive, quiescent hosts at cosmic noon. We summarize and quantify them below, in order of importance.

\begin{enumerate}
    \item \textit{Tidal evolution modeling.} The uncertainty in our current understanding of galaxy disruption is the largest systematic limiting our ability to make meaningful comparisons between simulations and observations of the galaxy population across cosmic time. Using a flexible, tidal track-based model for mass loss due to tidal stripping, we can vary the rate at which galaxies disrupt in our simulations. We find that changing galaxy durability, as described by the $f_{1/2}$ parameter, can alter satellite abundances by half an order of magnitude across the full range of $f_{1/2}$ values considered in this analysis. Even within the more physically motivated range of $0.001 \leq f_{1/2} \leq 0.1$, satellite abundances differ by as much as a factor of $\sim2$ at all redshifts and masses probed in our analysis. While current observational uncertainties remain too large to pin down the exact value and redshift evolution of $f_{1/2}$, our work suggests that satellites are at least as durable as hydrodynamic simulations predict, and may be $\gtrsim$2--3 times more durable than is typically assumed.

    \item \textit{Numerical disruption.} Once enough of a low-mass subhalo has been stripped away by tides, numerically induced runaway mass loss can destroy the subhalo. We implement an orphan model to approximate the true mass of the subhalo past the point where its mass loss rate is unconverged, and track it by its most-bound particle. We verified that the orbit and mass loss prescriptions in this orphan model can predict subhalos' true positions and masses with minimal bias; thus, this is a well-controlled source of uncertainty. For subhalos with peak particle counts of $n_{\mathrm{peak}} \lesssim 10^4$ (i.e., lower resolutions), implementing the orphan model increases satellite counts by a factor of $\sim1.8$. At high resolutions ($n_{\mathrm{peak}} \gtrsim 10^5$), there is no effect. 
    
    \item \textit{Host distribution mismatch.} The mass-selected SymphonyGroup zoom-in simulations result in a biased distribution of host halos (Figure \ref{fig:mass_distribution_shift}), and will thus lead to biased predictions about the satellite mass function. We leverage the self-similarity of subhalo mass functions at different host mass scales to approximate the satellite stellar mass function for a full, representative sample of host halos by rescaling the mass function of the SymphonyGroup halos (see Appendix \ref{app:mass distribution}). This primarily affects our two lower-redshift bins, where the SymphonyGroup host halo mass distribution is narrowest, altering satellite abundances by up to $\sim 50\%$. However, the subhalo/satellite mass function is not \textit{perfectly} self-similar with host mass \citep[e.g.,][]{Nadler2023}, especially at the higher end of the subhalo mass range, where host-to-host scatter can be significant (a factor of 2 or more; see Figure 7 in \citealt{Nadler2023}). Additional simulations spanning a wider range of host halo mass ($\sim10^{11.5-14}~\Msun$) should be run to fully and accurately capture the subhalo distribution around a population of hosts that match the available JWST observations.

    \item \textit{Baryon-induced halo contraction.} The stars embedded within dark matter halos increase the halos' central densities and can boost subhalo abundances within a given radius. We model this effect post-hoc using the adiabatic contraction prescription of \citet{Blumenthal1986} and find that our subhalo counts increase by $\lesssim 5\%$ at all redshifts probed (see Appendix \ref{app:adiabatic contraction}). Thus, this is a small, well-controlled effect.

    \item \textit{Interlopers from correlated structure.} We include the impact of large-scale structure on our subhalo abundances by modeling the expected number of unassociated subhalos that are projected into our search aperture from the correlated structure around massive host halos. This effect turns out to be straightforwardly modeled and small within our chosen 35 kpc aperture, $\ll 1\%$ across the $1.0 < z < 3.5$ redshift range (see Appendix \ref{app:interloper modeling}).
\end{enumerate}

To improve our theoretical predictions, we must address the limitations in our simulations. As alluded to previously, a central priority for upcoming work is to fill the host mass gap in our zoom-in simulations, particularly at $z \sim 1$. We plan to run a suite of high-resolution simulations spanning the $10^{11.5}$–$10^{14}~\Msun$ halo mass range to better capture the halo population corresponding to our observed host galaxies. 

Additionally, the host galaxy's baryonic potential can accelerate subhalo mass loss and meaningfully alter the predicted abundance and radial distribution of satellite galaxies \citep[e.g.,][]{BrooksZolotov2014, Sawala2017, Barry2023}. We do not model the tidal forces from the hosts' baryonic mass content in the present analysis. In future work, we will embed central galaxy potentials into host halos to account for enhanced tidal disruption due to baryons \citep[e.g.,][]{GarrisonKimmel2017, EDEN2024}. 

We will also improve our tidal evolution models by including the impact of subhalo internal structure and kinematics, as both the stellar profile and velocity anisotropy at infall of a subhalo can significantly influence its tidal track \citep[e.g.,][]{Errani2022, Chiang2024}. Accurately capturing these effects is essential for a robust theoretical framework linking dark matter and stellar mass loss in satellite galaxies.

\subsection{Uncertainty in the galaxy--halo connection}
\label{sec:GHC}

Additional systematics may be introduced by the specific choices we make in our galaxy--halo connection model. There are two important aspects of this modeling to consider: 1) uncertainty within the UM-DR1 model used in our work, and 2) extrapolations of that model applied to our low-mass, high-redshift regime.

The impact of the uncertainty in the UM-DR1 model parameters can be directly quantified. We measure the simulated satellite stellar mass functions while marginalizing over the UM-DR1 model posteriors, and find that this introduces a $\lesssim$10--15$\%$ variation in the predicted satellite SMFs. However, we note that our data probes a regime outside the range where UM-DR1 was constrained: the stellar mass functions used for the global UM-DR1 SHMR do not extend below $\sim10^{8.5}~\Msun$ at $z = 1$ and $\sim 10^{11}~\Msun$ at $z = 3$ \citep{UM2019}. The redshift range probed in our work also falls below the regime where UV luminosity functions begin to inform SHMR constraints at $z \gtrsim 4$. Thus, our model predictions for the low-mass satellite population at $1 < z < 3.5$ push the theoretical galaxy--halo connection into an extrapolated regime where it is challenging to robustly estimate the uncertainty introduced by the SHMR. In summary, within the context of a known stellar mass function, we expect that the galaxy--halo uncertainties are likely to be subdominant compared to other modeling uncertainties (see Section \ref{sec:theory systematics}), primarily because the dominant source of uncertainty in the \textsc{UniverseMachine} model stems from its treatment of galaxy disruption and orphan subhalos, which we have replaced and handled separately in our analysis. However, a full quantification of these uncertainties requires revisiting the faint end of the global stellar mass function at these epochs. In future work, we plan to address the issue of extrapolation by fitting a new set of \textsc{UniverseMachine} parameters to JWST data as part of a joint constraint with $f_{1/2}$.

In the \textsc{UniverseMachine} orphan model, orphans are removed once their modeled subhalo mass falls below a certain threshold \citep[see Appendix B of][]{UM2019}. This threshold lacks a direct connection to physical disruption mechanisms and is thus difficult to constrain. In contrast, our approach avoids an abrupt disruption criteria. Orphan subhalos are never removed; instead, we allow them to continue evolving under tidal forces, losing mass over time. If and when their stellar mass drops below the observational detection threshold, they cease to be counted. Using this physically motivated model of subhalo tidal evolution, we are able to effectively marginalize over satellite disruption physics and capture the uncertainty associated with subhalo durability. 

The assumed stellar profile of satellites also influences how rapidly its stars are stripped relative to its dark matter and the shape of its tidal track (see Figure \ref{fig:sersic_tidal_tracks}). In the case of Figure \ref{fig:nsat_vs_f0}, this can change the range of $f_{1/2}$ values compatible with observations at the $\sim60\%$ level. While less significant compared to other systematics, it can nonetheless meaningfully influence the quantitative interpretation of constraints on $f_{1/2}$.

%The extrapolation of the stellar mass--halo mass relation in this regime is particularly important for this work. Because we scale the SymphonyGroup subhalo masses to approximate a representative host sample, we must adopt a SHMR to convert those rescaled subhalo masses into stellar masses. In this analysis, we apply the best-fit UM-DR1 satellite SHMR \citep{UM2019}. However, the SymphonyGroup satellites tend to have systematically higher stellar masses than predicted by this median relation. As a result, applying the standard SHMR with an assumed 0.3 dex scatter leads to a systematic underestimation of satellite stellar masses. To avoid this, we preserve the SymphonyGroup satellites’ original offsets from the median relation. This discrepancy is not unexpected, given that our sample focuses on massive, quiescent host galaxies that likely reside in more biased environments than those used to calibrate UM-DR1. Nevertheless, it highlights that the SHMR can significantly impact the inferred stellar masses of our modeled satellites.
%However, because we are applying an extrapolation of the SHMR, its accuracy in this regime is unknown. 

\subsection{Observational systematics}
\label{sec:obs systematics}

This analysis is also subject to several observational systematics. First, satellites in the \citet{Suess2023} sample are identified based on photometric redshift consistency within $1\sigma$ of their hosts; the photometric redshift errors in this regime are not well calibrated.
A second concern is contamination from unassociated foreground or background sources that pass the photometric redshift selection (``random interlopers''). While our model accounts for projection effects from large-scale structure (``correlated interlopers''), it does not include contamination from random field alignments. \citet{Suess2023} estimated that up to 20\% of companions may be chance alignments, suggesting satellite abundances could be biased high at that level, though this fraction likely varies with redshift and stellar mass. Lastly, we note that our mass-completeness thresholds --- 1.5 dex below HST’s 90\% completeness limits --- are conservative given that JADES imaging is up to 2 magnitudes deeper than HST at comparable rest-frame wavelengths \citep{JADES2023}; a more precise characterization of completeness would enable us to truly leverage this deeper data and strengthen constraints on galaxy evolution models. 

In this work, we have quantified the dominant theory systematics at play in our analysis. In upcoming work, we will do the same for the observations. The statistical error budget will be minimized with the inclusion of more data covering a larger sky area. While \citet{Suess2023} primarily relied on observations from JADES-GOODS-S, multiple other portions of the CANDELS extragalactic fields have been observed with multi-band JWST imaging to $m_{\mathrm{f444w}}<28$ (e.g., CEERS; \citealt{CEERS2025} and PRIMER; \citealt{PRIMER2021}). Using all the publicly available JADES, CEERS, and PRIMER mosaics will result in a $\sim 8\times$ increase in area and an expected $\sim65\%$ decrease in the Poisson uncertainty on observed satellite mass functions. In combination with improved estimates of completeness, contamination, and photometric redshifts from forthcoming medium-band surveys in this field (MINERVA, PID 7814; SPAM, PID 8559), these data will significantly improve the reliability of observational constraints on the satellite population.

%%%%%%%%%%%%%%%%%%%%%%%%%%%%%%%%%%%%%%%%%%%%%%%%%%%%%%%%%%%%%%%%%%%%%%%%%%

\section{Discussion \& Conclusions} \label{sec:conclusions}

In this paper, we build a robust theoretical model that enables direct, one-to-one comparisons between simulated and observed satellite populations. Using this model, we evaluate the stellar mass functions and overall abundances of satellites within 35 kpc of massive, quiescent galaxies at redshifts $1 < z < 3.5$ using JWST data \citep{Suess2023} and cosmological dark matter-only zoom-in simulations (SymphonyGroup; \citealt{Symphony}).

Critically, we quantify and account for several key systematics that affect theoretical satellite abundances:
\begin{enumerate}
    \item \textit{Tidal evolution modeling.} Satellite counts can vary by a factor of $\sim1.5$--$3.5$ depending on the assumed durability of satellite galaxies against tidal stripping.
    
    \item \textit{Numerical disruption of halos.} Approximating the masses and positions of subhalos that are no longer numerically reliable using an orphan model increases satellite counts by a factor of $\sim1.8$ at lower subhalo resolutions ($n_{\mathrm{peak}} \lesssim 10^4$).
     
    \item \textit{Host distribution mismatch.} The underlying distribution shift between the observed and simulated host samples can lead to $\sim 50\%$ discrepancies in their respective satellite abundances if not corrected for; this shift is expected to be $\lesssim10-20\%$ when an appropriate correction is applied.

    %\item \textit{The galaxy--halo connection.} Predicted satellite counts are sensitive to the global galaxy--halo connection. Marginalizing over the UM-DR1 posterior introduces a $\lesssim$10--15$\%$ variation in the predicted satellite SMFs; fully accounting for the uncertain extrapolation to lower masses will require further work on the field stellar mass function.

    \item \textit{Baryon-induced halo contraction.} The increase in host central density due to the presence of baryons leads to a $\lesssim 5\%$ increase in satellite counts within the inner region of the host.

    \item \textit{Interlopers from correlated structure.} The projection of galaxies associated with large-scale structure around host halos contributes $\lesssim1\%$ to our predicted satellite abundances.  
\end{enumerate}

Predicted satellite counts are sensitive to the global galaxy--halo connection. Marginalizing over the UM-DR1 posterior introduces a $\lesssim$10--15$\%$ variation in the predicted satellite SMFs; fully accounting for the uncertain extrapolation to lower masses will require further work on the field stellar mass function.

Baryonic effects can also significantly contribute to disruption of the satellite population through feedback processes, dominated by the impact of the central galaxy \citep[e.g.,][]{Chan2015, GarrisonKimmel2017, Sales2022, EDEN2024}. However, to isolate the impact of baryonic feedback, we must first establish a reliable baseline for how substructure evolves in its absence. In this sense, gravitational tides represent the minimal physical process governing subhalo survival. 
%RW I am not sure if this sentence is needed here.  I put some of the words above....
%If we lack a clear understanding of how gravitational tides strip dark matter and stars from subhalos after infall, then theoretical predictions risk being systematically biased and potentially misleading.

A key advance in this work lies in the implementation of a flexible model for how gravitational tides strip dark matter and stars from subhalos after infall, an essential element to robust theoretical predictions.  This tidal track-based model is directly integrated into our subhalo orphan model, and allows us to explore the effect of galaxy durability on the predicted satellite population as well as to constrain the range of galaxy tidal evolution models that are consistent with observations. By comparing our modeled satellite abundances as a function of galaxy durability, quantified as $f_{1/2}$, with observational measurements, we conclude that simulated satellites are likely at least as durable as that predicted by hydrodynamic simulations, and possibly more durable at redshifts $z > 1.2$. The observations may suggest a stronger redshift evolution than our model allows; however, additional work including a larger dataset is necessary to draw a robust conclusion about redshift evolution.

A wide range of astrophysical applications --- including in galaxy formation, cosmological inference, and dark matter physics --- critically depend on how well we understand the durability of subhalos in cosmological simulations. In galaxy formation modeling, uncertainties in subhalo durability propagate directly into the inferred galaxy--halo connection, and can lead to biases in satellite population predictions and galaxy clustering statistics, especially on small scales \citep[e.g.,][]{Yang2012, Reddick2014, Lim2017, Pujol2017, Campbell2018}. Furthermore, accurate modeling of subhalo survival rates is essential for predicting the small-scale matter power spectrum and the growth of structure, which in turn affects cosmological parameter constraints from observables such as galaxy clustering and weak lensing \citep[e.g.,][]{Trujillo-Gomez2011, Lange2023}. The abundance and spatial distribution of subhalos are also key observables for discriminating between different dark matter models \citep[e.g.,][]{Spergel2000, Bode2001, TulinYu2018, Nadler2021}, but our ability to make accurate predictions about the properties of substructure hinges on robust modeling of subhalo tidal evolution. 

Bridging this theoretical gap requires both improved modeling and high-quality data capable of testing those models. As this study demonstrates, new data from JWST enables strong empirical constraints on the evolution of satellite systems over a broad cosmic timescale. Our simulation-based forward modeling approach, including advances in subhalo tracking and modeling, fully enables this. By systematically quantifying and incorporating the dominant sources of uncertainty, here we establish a flexible framework for interpreting high-redshift satellite abundances and for constraining the physics of galaxy disruption and the galaxy--halo connection. This framework can be applied to a wide range of data over mass scales and epoch, from JWST to the Milky Way to the large volumes probed by cosmic surveys. As these instruments continue to deliver deeper and more extensive datasets, and as new simulations broaden the mass and physical scope of our models, we will be poised to decode the tidal histories of satellite galaxies across cosmic time.

%%%%%%%%%%%%%%%%%%%%%%%%%%%%%%%%%%%%%%%%%%%%%%%%%%%%%%%%%%%%%%%%%%%%%%%%%
%TC:ignore
\begin{acknowledgments}
We thank Ethan Nadler and Tara Dacunha for helpful comments on a draft.
JTW was supported in part by an NSF Graduate Research Fellowship. This work is a part of program JWST-AR-05907; support for this program was provided by NASA through a grant from the Space Telescope Science Institute, which is operated by the Association of Universities for Research in Astronomy, Inc., under NASA contract NAS 5-03127. KS acknowledges support provided by NASA through Hubble Fellowship grant HST-HF2-51543.001-A awarded by the Space Telescope Science Institute, which is operated by the Association of Universities for Research in Astronomy, Inc., for NASA, under contract NAS 5-26555. Additional support was provided by the Kavli Institute for Particle Astrophysics and Cosmology at Stanford and the SLAC National Accelerator Laboratory. The work of CCW is supported by NOIRLab, which is managed by the Association of Universities for Research in Astronomy (AURA) under a cooperative agreement with the National Science Foundation.  

This work is based in part on observations made with the NASA/ESA/CSA James Webb Space Telescope. The data were obtained from the Mikulski Archive for Space Telescopes at the Space Telescope Science Institute, which is operated by the Association of Universities for Research in Astronomy, Inc., under NASA contract NAS 5-03127 for JWST. These observations are associated with the JADES (\citealt{JADES2023},
PID 1180), JEMS (\citealt{JEMS2023}, PID 1963), and FRESCO
(\citealt{FRESCO2023}, PID 1895) programs.

This work used data from the Symphony suite of simulations ({\tt http://web.stanford.edu/group/gfc/symphony/}), 
which was supported by the Kavli Institute for Particle Astrophysics and Cosmology at Stanford University and SLAC National Accelerator Laboratory, and by the U.S. Department of Energy under contract number DE-AC02-76SF00515 to SLAC National Accelerator Laboratory. We thank all of our Symphony collaborators for their contributions to this enabling work.

This research made use of computational resources at SLAC National Accelerator Laboratory, a U.S.\ Department of Energy Office; the authors are thankful for the support of the SLAC computational team.

\end{acknowledgments}

%%%%%%%%%%%%%%%%%%%%%%%%%%%%%%%%%%%%%%%%%%%%%%%%%%%%%%%%%%%%%%%%%%%%%%%%

%TC:ignore
\appendix

\section{Correcting the host mass distribution mismatch between observations and SymphonyGroup} 
\renewcommand\thefigure{A\arabic{figure}}
\setcounter{figure}{0}
\label{app:mass distribution}

\begin{figure*}
    \centering
    \includegraphics[width=0.98\textwidth]{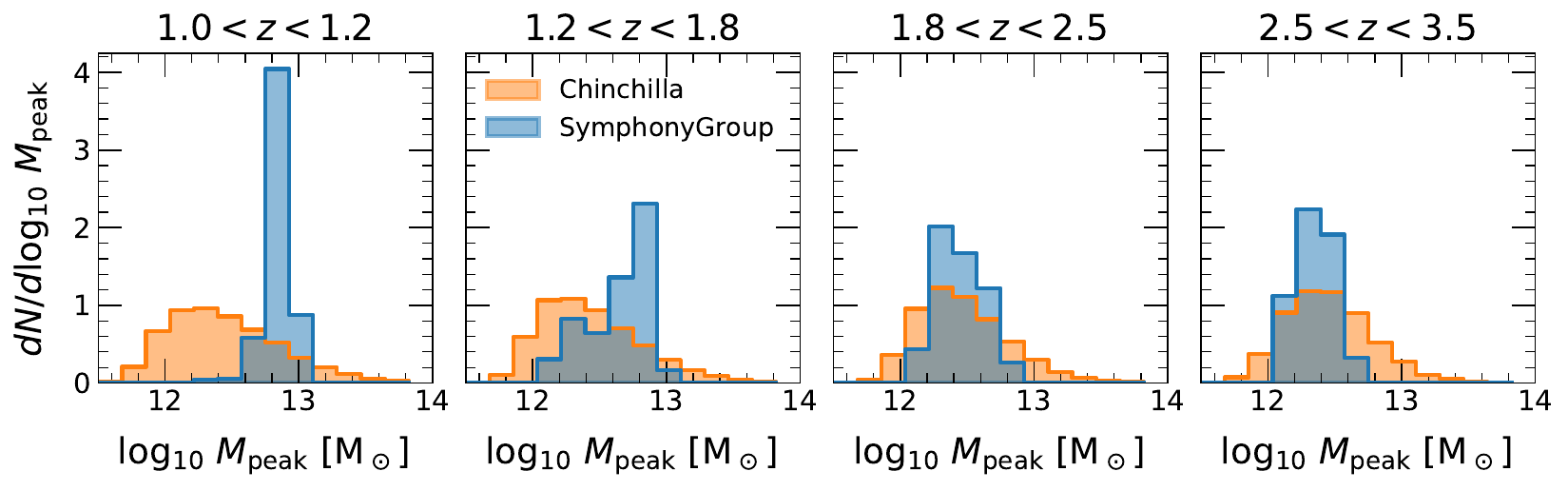}
    \caption{Distribution of SymphonyGroup host historical peak halo masses ($\Mpeak$; blue histograms) compared to the distribution of host peak halo masses from the Chinchilla Lb250-2048 cosmological box (orange histograms) using the same host selection criteria (stellar mass $\geq 10^{10}~\Msun$, sSFR $\leq 10^{-10.5}~\Msun$ yr$^{-1}$), in four bins of redshift between $1 < z < 3.5$. The SymphonyGroup host halos are clearly not a representative sample of all the halos that could host massive, quiescent galaxies in the redshift range of interest. The $\Mpeak$ distributions of the SymphonyGroup hosts are biased high relative to the Chinchilla hosts in the lower two redshift bins and under-sample the high-mass end in the highest redshift bin.}
    \label{fig:mass_distribution_shift}
\end{figure*}

The Symphony simulation suite that most closely matches the massive, quiescent host galaxies observed in \citet{Suess2023} is the SymphonyGroup zoom-in suite. The SymphonyGroup hosts were selected to have halo masses within the narrow range of $10^{12.86 \pm 0.10}~\Msun$ at $z = 0.5$, meaning that the SymphonyGroup host halos will not be a fully representative sample of \textit{all} the halos that could host the massive, quiescent (stellar mass $\geq 10^{10}~\Msun$, sSFR $\leq 10^{-10.5}~\Msun$ yr$^{-1}$) galaxies in the redshift range $1 < z < 3.5$ observed by \citet{Suess2023}. This is especially true at redshifts close to $z = 0.5$, where by construction, the SymphonyGroup mass distribution approaches a delta function.

This problem is clearly reflected in Figure \ref{fig:mass_distribution_shift}. Here, we plot the ``historical'' peak halo mass\footnote{The historical peak mass is the maximum of $\Mvir$ achieved by a halo up to the point of observation, \textit{not} the peak $\Mvir$ that the halo will ever achieve.} ($\Mpeak$) distributions of the SymphonyGroup hosts in four redshift bins. We compare this to the $\Mpeak$ distribution of hosts in the ``Chinchilla'' Lb250-2048 cosmological box, which is a dark matter-only simulation run with \textsc{L-Gadget}, a gravity-only version of Gadget-2 \citep{Springel2005}, that has $2048^3$ particles and a side length of 250 Mpc $h^{-1}$ \citep{Lehmann2017}. We identified host halos in Chinchilla using the same mass and sSFR selection criteria as mentioned above.

At most redshifts, there is a significant difference between the $\Mpeak$ distributions of the SymphonyGroup hosts and the Chinchilla hosts. This is especially true in the lowest-redshift bin, where the SymphonyGroup simulations fail to capture the population of $\sim 10^{12}~\Msun$ halos that can host the massive, quiescent galaxies seen in the observations. Overall, it is plain that the $\Mpeak$ distributions of the SymphonyGroup hosts are biased high relative to the Chinchilla hosts in the lower two redshift bins and undersample the high-mass end in the higher redshift bins. The abundance of subhalos above a fixed mass within a dark matter halo depends primarily on the mass of the halo \citep[e.g.,][]{Kravtsov2004}. Therefore, having an unbiased sample of host halo masses is crucial for making accurate predictions about the satellite mass function. We note that there is a secondary dependence of subhalo abundance on host halo concentration \citep{Zentner2005, Mao2015}. We have verified that, once the SymphonyGroup host mass distribution matches that of Chinchilla, the SymphonyGroup halo concentration distribution also matches Chinchilla.

We cannot, however, simply measure the satellite mass functions using the Chinchilla box because its resolution is much lower than that of our zoom-ins --- we would be unable to resolve subhalos down to the satellite masses probed by the JWST observations. In principle, the best solution would be to run a series of zoom-in simulations that span the full mass range of interest ($\approx 10^{11.5 - 14}~\Msun$), which we plan to do for future studies. At the time of this work, the next-closest existing zoom-in suites in mass are SymphonyMilkyWay (selected to have $\Mvir = 10^{12.09 \pm 0.02}~\Msun$ at $z = 0$) and SymphonyL-Cluster ($\Mvir = 10^{14.62 \pm 0.11}~\Msun$ at $z = 0$). At redshifts of $1 < z < 3.5$, the Milky Way hosts have masses that fall below the requisite mass range, while the L-Cluster hosts have masses that exceed it; thus, we must turn to alternative methods based on rescaling.

\begin{figure*}
    \centering
    \includegraphics[width=0.98\textwidth]{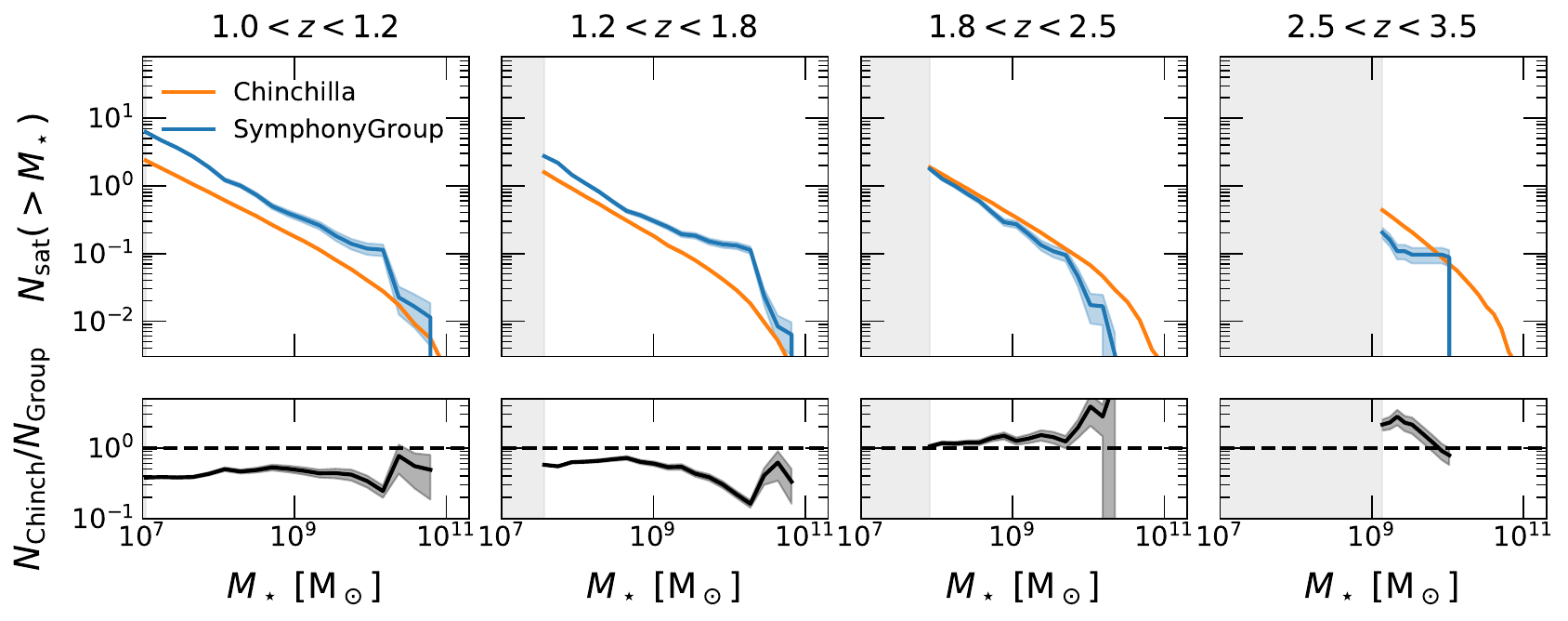}
    \caption{Impact of host halo mass distribution on the measured satellite abundance. The top row shows the average cumulative satellite mass function within a 2D projected radius of 35 kpc as measured from the SymphonyGroup host halos (blue lines) in four redshift bins. The blue shaded regions indicate their associated $1\sigma$ Poisson errors. The orange curves show the cumulative mass functions estimated for the Chinchilla hosts using the procedure described in this section.
    The light gray vertical bands indicate the mass ranges below the observational mass-completeness limits. The bottom row plots the ratio between the Chinchilla and SymphonyGroup mass functions. In the $1 < z < 1.2$ and $1.2 < z < 1.8$ bins, the satellite counts estimated from Chinchilla are systematically lower (by $\sim 50\%$) than what is measured in the SymphonyGroup zoom-ins. In the $1.8 < z < 2.5$ and $2.5 < z < 3.5$ bins, the overall normalization of the satellite mass function is generally consistent between the SymphonyGroup and Chinchilla hosts; however, there is a higher abundance of massive satellites estimated around the Chinchilla hosts.}
    \label{fig:scaled_mass_functions}
\end{figure*}

$N$-body simulations have shown that the form of the subhalo mass function (SHMF) is nearly self-similar when measured at a given redshift and host virial radius fraction $r/\Rvir$, with its amplitude scaling linearly with host mass (\citealt{Giocoli2010, Jiang2016, BBK2017}; see also Figures 7 and 8 of \citealt{Symphony}). We tested this self-similarity by rescaling the SymphonyGroup SHMFs according to the host mass ratio between SymphonyGroup and SymphonyMilkyWay to estimate the SymphonyMilkyWay SHMFs, finding agreement at the $\sim10-20\%$ level across all subhalo masses, well within the host-to-host scatter. We also verified that this self-similarity holds when $r$ is a 2D projected radius and subhalo mass is translated into stellar mass. We can exploit this property to approximate the subhalo --- and therefore, satellite --- abundances of host halos at a range of masses using just the SymphonyGroup halos.

In our analysis, we correct for the distribution mismatch between the observed and simulated SymphonyGroup hosts as follows. In a given redshift bin, we sample 10,000 host halos from the Chinchilla host distribution shown in Figure \ref{fig:mass_distribution_shift}. These hosts are characterized by their historical peak masses, $M_{\rm{peak}}^{\rm{C}}$, and virial radii, $R_{\rm{vir}}^{\rm{C}}$. Each Chinchilla host is then paired with a randomly selected SymphonyGroup host ($M_{\rm{peak}}^{\rm{SG}}$; $R_{\rm{vir}}^{\rm{SG}}$) from the same redshift bin, whose subhalo population is used as a proxy for the Chinchilla host’s subhalo population. Assuming that the amplitude of the subhalo mass function is linear in host mass, we estimate the masses of the Chinchilla subhalos by scaling the masses of the SymphonyGroup subhalos by $M_{\rm{peak}}^{\rm{C}}/M_{\rm{peak}}^{\rm{SG}}$. Additionally, we only consider SymphonyGroup subhalos that are within a projected radius of $r = 35~\mathrm{kpc} \times R_{\rm{vir}}^{\rm{G}}/R_{\rm{vir}}^{\rm{C}}$. This ensures that we are approximating the Chinchilla subhalo distribution within a fixed 35 kpc aperture, matching the observations. The infall stellar masses of the Chinchilla subhalos are determined using the median observed satellite stellar mass--halo mass relation from in \citet[][Appendix J]{UM2019}, with each subhalo’s offset from the median relation preserved based on its original offset in the SymphonyGroup simulation. Satellite stellar masses at later times are calculated using the tidal evolution model described in Section \ref{sec:subhalo mass loss}. The final satellite stellar mass function is then the average over the 10,000 Chinchilla hosts, weighted such that the stellar mass distribution of the Chinchilla hosts matches that of the observed hosts in the chosen redshift bin. This procedure yields the predicted satellite stellar mass functions that we use throughout the main analysis.

Figure \ref{fig:scaled_mass_functions} compares the satellite mass functions measured from the SymphonyGroup zoom-ins against those estimated for an unbiased host sample from Chinchilla. The top row of the figure shows the average cumulative satellite stellar mass functions measured from the SymphonyGroup host halos and approximated for the Chinchilla hosts in four redshift bins. The bottom row shows the ratio between the Chinchilla and SymphonyGroup satellite abundances as a function of satellite mass.

We find that accounting for the incorrect mass distribution of the SymphonyGroup host halos in the $1 < z < 1.2$ and $1.2 < z < 1.8$ bins results in a $\gtrsim 50\%$ suppression in satellite abundances at all masses. This is consistent with the SymphonyGroup host mass distributions being biased high relative to the ``true'' Chinchilla host mass distribution in these bins. In the $1.8 < z < 2.5$ and $2.5 < z < 3.5$ bins, the overall normalization of the satellite mass functions remain more or less unchanged. This is expected, as the median $\Mpeak$ values of the SymphonyGroup and Chinchilla hosts are consistent with one another at these redshifts. However, the Chinchilla host mass distribution includes a high-mass tail that is not sampled by the SymphonyGroup zoom-ins. This results in the excess of high-mass satellites seen in the Chinchilla satellite stellar mass functions relative to that of SymphonyGroup.

Altogether, the incorrect mass distribution of the SymphonyGroup host halos can alter predicted subhalo abundances by more than $\sim50\%$ at $z \sim 1$. The size of this effect is substantial enough to influence conclusions made about the consistency between observed and predicted satellite distributions, which emphasizes the need to fill in the mass gaps between the Symphony zoom-in suites. As the correction procedure described in this section underlies the satellite mass functions used throughout our analysis, properly accounting for the host mass distribution is critical for robust comparison to observations.

%%%%%%%%%%%%%%%%%%%%%%%%%%%%%%%%%%%%%%%%%%%%%%%%%%%%%%%%%%%%%%%%%%%%%%%%%

\section{Accounting for the effect of baryons on host halo central densities} \label{app:adiabatic contraction}
\renewcommand\thefigure{B\arabic{figure}}
\setcounter{figure}{0}

The gravitational field of a galaxy in the center of a dark matter halo contracts the halo's dark matter at smaller radii. This effect steepens the halo's central density profile and boosts the number of subhalos within a given radius. As our simulations are dark matter-only, this effect will bias our subhalo mass functions low. We account for the expected enhancement in subhalo counts using the standard adiabatic contraction approach \citep{Blumenthal1986, Gnedin2004}.

Assuming spherical symmetry, circular particle orbits, conservation of angular momentum, and no shell-crossing, the final dark matter distribution can be calculated using
\begin{equation}
    \big[M_{\rm{DM}}(r_i) + M_b(r_i)\big]r_i = \big[M_{\rm{DM}}(r_f) + M_b(r_f)\big]r_f,
    \label{eq:adiabatic contraction}
\end{equation}
given the initial dark matter and baryon mass profiles $M_{\rm{DM}}(r_i)$ and $M_b(r_i)$ and final baryon profile $M_b(r_f)$. 

\begin{figure*}
    \centering
    \includegraphics[width=0.9\textwidth]{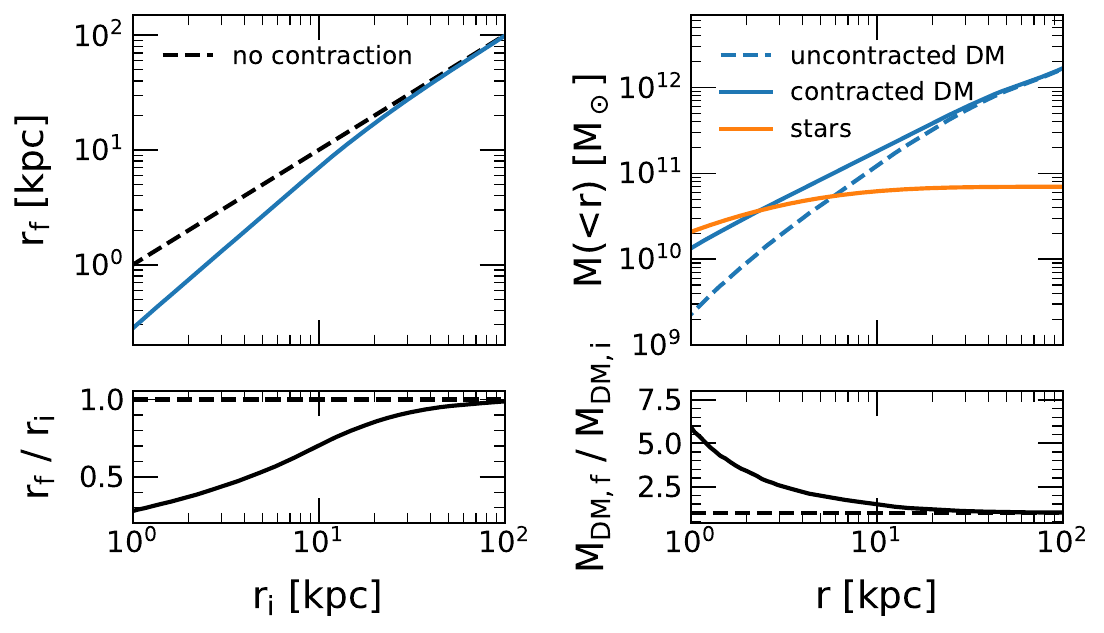}
    \caption{An example of the effect of adiabatic contraction on a $10^{12.4}~\Msun$ host halo at $z = 2$ in which a galaxy of stellar mass $10^{11}~\Msun$ resides. \textit{Left}: The top panel shows the relationship between the initial, uncontracted radii ($r_i$) and the final, contracted radii ($r_f$) of the halo. The black dashed line denotes the $\mathrm{r_f = r_i}$ line (i.e., no contraction). The lower panel shows the ratio between $r_f$ and $r_i$ as a function of $r_i$. \textit{Right}: The top panel compares the uncontracted (dashed blue line) and contracted (solid blue line) halo mass profiles. The assumed final stellar mass profile is also plotted in orange. The bottom panel shows the ratio between the final and initial enclosed halo masses as a function of radius. Particles within a radius of $r_i \approx 38$ kpc are contracted down to $r_f \approx 35$ kpc (the satellite search aperture used in our analysis), leading to a $\sim8\%$ increase in enclosed dark matter mass within 35 kpc. The effect size drops off with radius, such that there is essentially no radial contraction at the virial radius and beyond.}
    \label{fig:contracted_profiles}
\end{figure*}

The assumption of no shell-crossing enforces that in the process of contraction, no layer --- or ``shell'' --- of dark matter intersects another layer. This means that within a sphere of initial radius $r_i$ that is contracted to $r_f$, the enclosed dark matter mass remains the same. In other words, $M_{\rm{DM}}(r_i) = M_{\rm{DM}}(r_f)$. Furthermore, assuming that the initial baryon profile follows that of the dark matter with some universal baryon fraction $f_b$, i.e., $M_b(r_i) = f_b \cdot M_{\rm{DM}}(r_i)$, the adiabatic contraction equation simplifies to
\begin{equation}
    (1 + f_b)~M_{\rm{DM}}(r_i)~r_i = \big[ M_{\rm{DM}}(r_i) + M_b(r_f) \big]r_f.
\end{equation}
Here, the initial dark matter profile $M_{\rm{DM}}(r_i)$ of a halo can be measured directly from the simulation particle data. The baryon fraction $f_b$ can be calculated using the catalog halo mass and the galaxy stellar mass estimated by \textsc{UniverseMachine}. Then, all that's left to do is to determine the shape of the final stellar profile. 

Because we selected for simulated host galaxies in this work to be massive ($M_* \geq 10^{10}~\Msun$) and quiescent (sSFR $\leq 10^{-10.5}$ yr$^{-1}$), we assume that they are elliptical galaxies and follow a S\'ersic profile with $n = 4$. The 2D S\'ersic profile can be deprojected into 3D with an Abel transform to obtain a circularized, three-dimensional mass profile:
\begin{equation}
    \rho_\star(r) = \frac{b_n}{\pi} \frac{\Sigma_0}{r_e} \Big( \frac{r}{r_e} \Big) ^{\frac{1}{n} - 1} \int_1^\infty \frac{\exp \big[ b_n (\frac{r}{r_e})^{1/n}~t \big]}{\sqrt{t^{2n} - 1}} dt,
    \label{eq:deproj_sersic}
\end{equation}
where $b_n \approx 2n - \frac{1}{3} + \frac{4}{405n} + \frac{46}{25515n^2}$ \citep{CiottiBertin1999}, $r_e$ is the 2D half-light radius, and $\Sigma_0$ is the normalization. We assume a constant mass-to-light ratio throughout the galaxy. 
%\footnote{In principle, as the S\'ersic model describes the surface brightness profile of a galaxy, the deprojection results in a three-dimensional \textit{light} profile. However, this is related to the mass density profile simply by a factor of mass-to-light ratio, which gets absorbed into the $\Sigma_0$ normalization term.} 
The size of a given host galaxy, $r_e$, is approximated using the size-mass relation from \citet{vanderWel2014}. The normalization, $\Sigma_0$, can then be calculated by requiring that the enclosed mass within $r = \infty$ is the total stellar mass of the galaxy. 

With that, we have all the pieces required to solve Equation \ref{eq:adiabatic contraction} for the relationship between $r_i$ and $r_f$, which we do for each relevant host halo and at each relevant snapshot. An example of this is shown in Figure \ref{fig:contracted_profiles} for a single host halo at $z = 2$. The left panel displays $r_f$ as a function of $r_i$, and the right panel plots the uncontracted (dashed blue line) and contracted (solid blue line) halo mass profiles. The assumed final stellar mass profile is also shown in orange. 

Because we search for satellites within a 2D projected radius of 35 kpc from their hosts in our main analysis, we will examine the impact that adiabatic contraction has on the host halo profile at 35 kpc as a case study. We find that particles initially within a radius of $r_i \approx 38$ kpc are pulled in to $r_f \approx 35$ kpc in the contracted halo. This translates to a $\sim8\%$ increase in enclosed dark matter mass at 35 kpc between the uncontracted and contracted halos. The effect size drops off with radius, such that there is no radial contraction at the virial radius and beyond.

\begin{figure}
    \centering
    \includegraphics[width=0.47\textwidth]{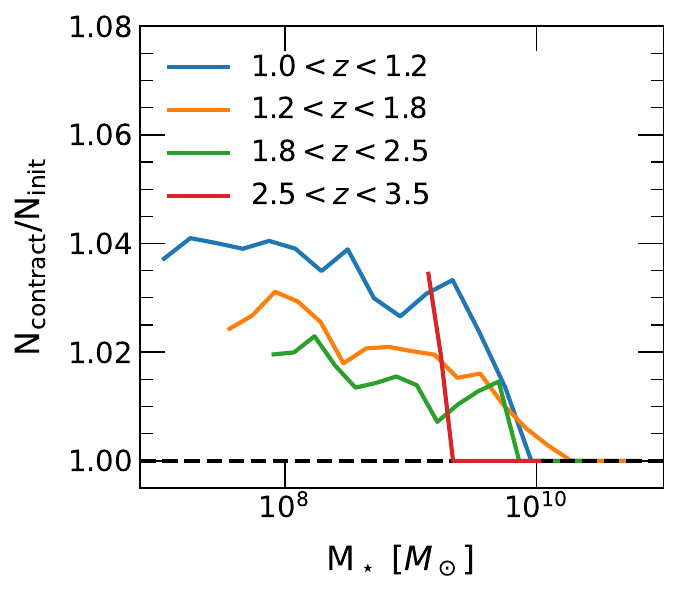}
    \caption{The ratio between the average per-host number of simulated satellites that fall within the selection criteria enumerated in Section \ref{sec:simulations} with and without accounting for adiabatic contraction of the host halo, as a function of stellar mass. The blue, orange, green, and red lines represent the four redshift bins used in our analysis: $1 < z < 1.2$, $1.2 < z < 1.8$, $1.8 < z < 2.5$, and $2.5 < z < 3.5$. We find that the effect size tends to decrease with redshift and is overall a $\lesssim 5\%$ effect at all satellite masses and redshifts.}
    \label{fig:contraction_effect_size}
\end{figure}

Assuming locations of subhalos contract in the same manner as the dark matter halo, we can adjust the positions of the subhalos accordingly --- a subhalo initially sitting at a radial separation of $r_i$ from the host now has a radial separation of $r_f$. In Figure \ref{fig:contraction_effect_size}, we plot the ratio between the average number of simulated satellites per host that fall within the selection criteria enumerated in Section \ref{sec:simulations} with and without accounting for adiabatic contraction of the host halo, as a function of stellar mass, in four bins of redshift. We find that the effect size is overall $\lesssim 5\%$ at all satellite masses and redshifts, and tends to decrease with redshift.

As shown in Figure \ref{fig:contracted_profiles}, it may seem like this should be a $\sim10\%$ effect rather than a few percent effect; however, this is not the case. The primary reason is that, to match observations, we are counting satellites in projection. The true 3D host--satellite separations for a large fraction of satellites within the 2D aperture are actually $\sim20$--$30\%$ larger than 35 kpc. This serves to wash out the impact of the contraction, as the bulk effect of adiabatic contraction drops off quickly with distance from the host center. 

%%%%%%%%%%%%%%%%%%%%%%%%%%%%%%%%%%%%%%%%%%%%%%%%%%%%%%%%%%%%%%%%%%%%%%%%%

\section{Adding in the contribution of interlopers from correlated structure} \label{app:interloper modeling}
\renewcommand\thefigure{C\arabic{figure}}
\setcounter{figure}{0}

The total number of satellites observed around a host galaxy is a combination of true satellites and background/foreground galaxies that happen to fall within the chosen aperture and photometric redshift range. These latter interloper galaxies are a combination of ``random'' galaxies and ``correlated'' galaxies. Random interlopers are spatially invariant across the sky, while correlated interlopers come from local large-scale structure that the host galaxy is embedded within. Quantifying the impact of random interlopers is achieved in observations by running the satellite detection pipeline on random sky patches \citep{Suess2023}. Here, we describe our approach to accounting for interlopers due to correlated structure in our theoretical predictions.

\begin{figure*}
    \centering
    \includegraphics[width=0.8\textwidth]{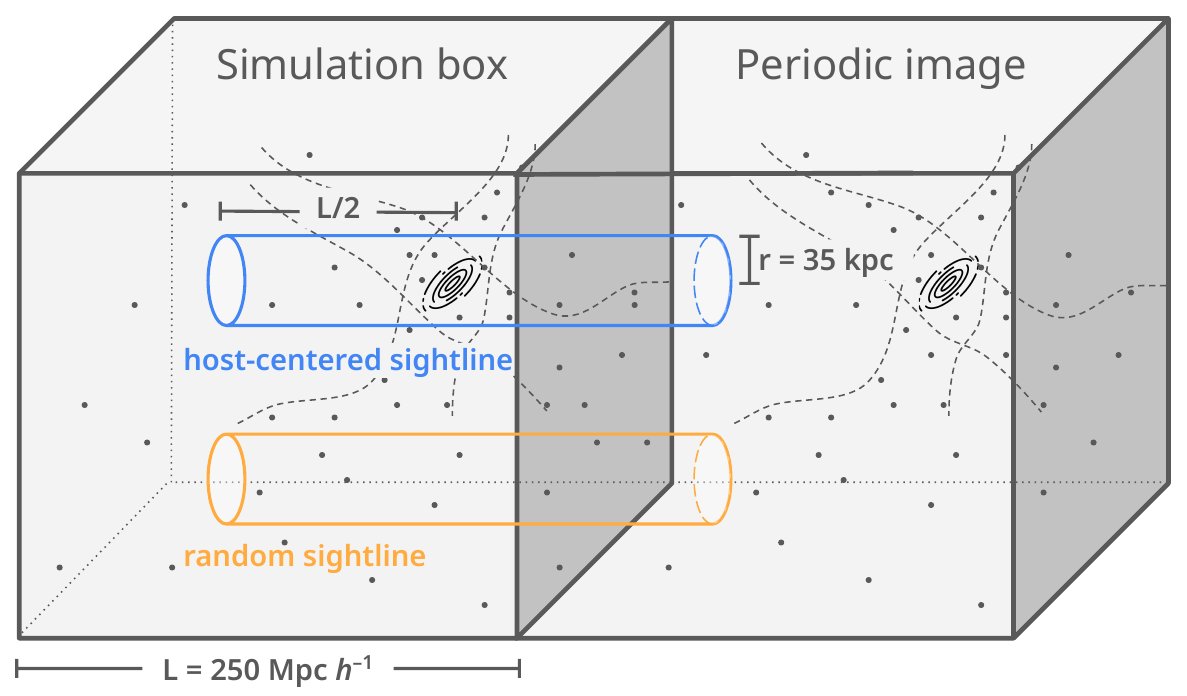}
    \caption{Cartoon depicting the procedure for counting interlopers along host-centered versus random sightlines. A ``host-centered sightline'' is defined as a cylinder of radius 35 kpc, whose central axis points along the $\hat{x}$ (or $\hat{y}$ or $\hat{z}$) direction, around a host galaxy of stellar mass $\geq 10^{10}~\Msun$. A ``random sightline'' is similarly defined around a random position in the simulation box that is further than 10 Mpc $h^{-1}$ from a host. To ensure that we will eventually move out of the correlated structure regime and into the field when counting galaxies along host-centered sightlines, we enforce periodic boundary conditions on the simulation box.}
    \label{fig:interlopers_cartoon}
\end{figure*}

When galaxies above a certain mass threshold are tallied as a function of line-of-sight distance from a massive host, the cumulative count initially grows rapidly in the overdense environment near the host. Beyond this ``correlated structure'' regime, the rate of increase will eventually become constant, with the total number of galaxies rising linearly with the average galaxy number density of the universe. That is to say, 
\begin{equation}
    \begin{aligned}[b]
        N_{\mathrm{tot}}(r) = & ~N_{\mathrm{sats}}(r < \Rvir) ~+ \\
        & ~\Nint(\Rvir < r < R_{\mathrm{random}}) ~+ \\
        & ~\Nrand(r > R_{\mathrm{random}}),
    \end{aligned}
\label{eq:N_satellites}
\end{equation}
where $N_{\rm{tot}}$ is the total number of galaxies above some given mass threshold that are at a LOS distance $\leq r$ from the host, $N_{\rm{sats}}$ is the satellite count of the host (i.e. the number of galaxies within the virial radius $\Rvir$ of the host), $\Nint$ is the number of galaxies within the correlated structure, $\Nrand$ is the number of field galaxies, and $R_{\rm{random}}$ is the distance from the host at which you hit the field. Using this general framework, we can quantify $\Nint$ as a function of interloper (stellar) mass threshold and redshift. We use the Chinchilla Lb250-2048 cosmological box with periodic boundary conditions (see Figure \ref{fig:interlopers_cartoon}) to carry out this interloper modeling. 

At each simulation snapshot between $1 < z < 3.5$, we randomly select 5000 hosts in the simulation box that have stellar masses above $10^{10}~\Msun$, consistent with the observational sample. A ``host-centered sightline'', i.e., a cylinder of radius 35 kpc whose central axis points along the $\hat{x}$ direction, is defined around each host. Starting at a distance $\Rvir$ from the host, we add up the number of galaxies as a function of separation from the host along each sightline, out to a maximum distance of 125 Mpc $h^{-1}$ (half the simulation box length). This is done for five different minimum stellar mass limits evenly spaced in the log between $10^7~\Msun$ and $10^9~\Msun$. This procedure allows us to measure $N_{\rm{host}} \equiv N_{\rm{tot}} - N_{\rm{sats}}$ in Equation \ref{eq:N_satellites}. 

Additionally, we define 5000 ``random sightlines'' centered on random locations in the simulation box and similarly count the number of galaxies above the same aforementioned stellar mass limits as a function of distance. These random sightlines probe $N_{\rm{random}}$ in Equation \ref{eq:N_satellites}. The difference between $N_{\rm{host}}$ and $N_{\rm{random}}$ then tells us the contribution of interlopers arising from correlated structure.

This procedure is repeated in both the $\hat{y}$ and $\hat{z}$ directions as well. All galaxy counts reported here are an average over the $\hat{x}$, $\hat{y}$, and $\hat{z}$ viewing orientations.

\begin{figure*}[!htbp]
    \centering
        \subfigure{\hspace{3mm}\includegraphics[width=.45\linewidth]{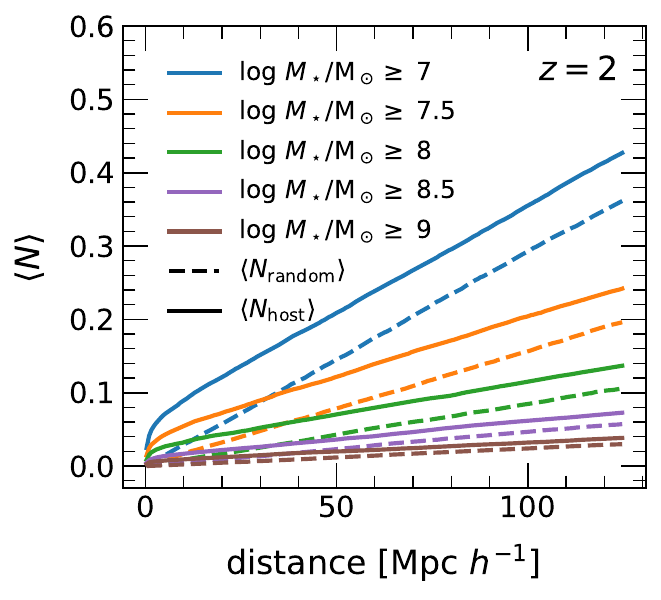}}\hfill
        \subfigure{\includegraphics[width=.465\linewidth]{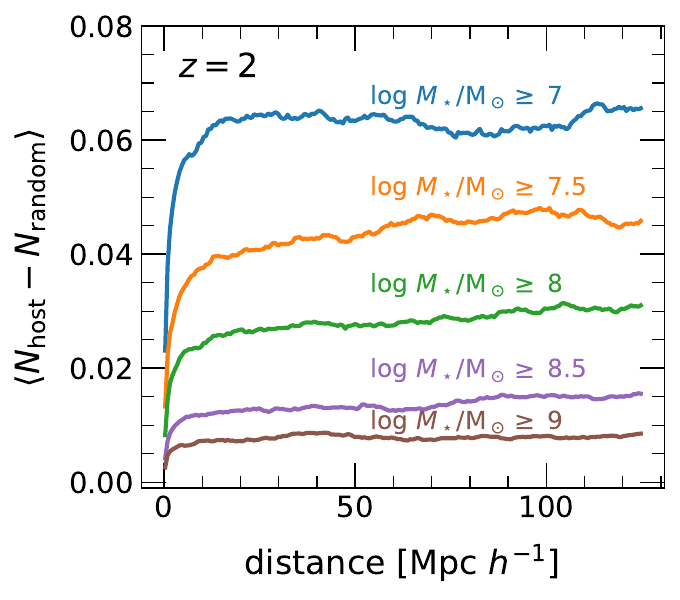}}\hfill
        \subfigure{\includegraphics[width=.475\linewidth]{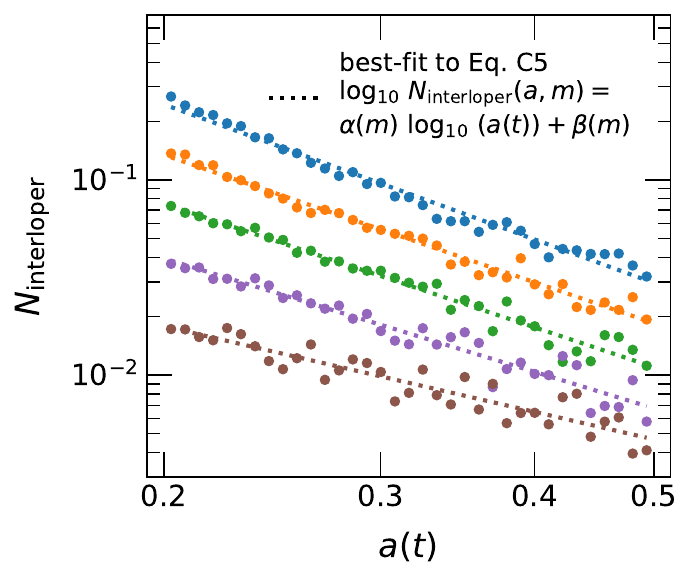}}\hfill
        \subfigure{\includegraphics[width=.485\linewidth]{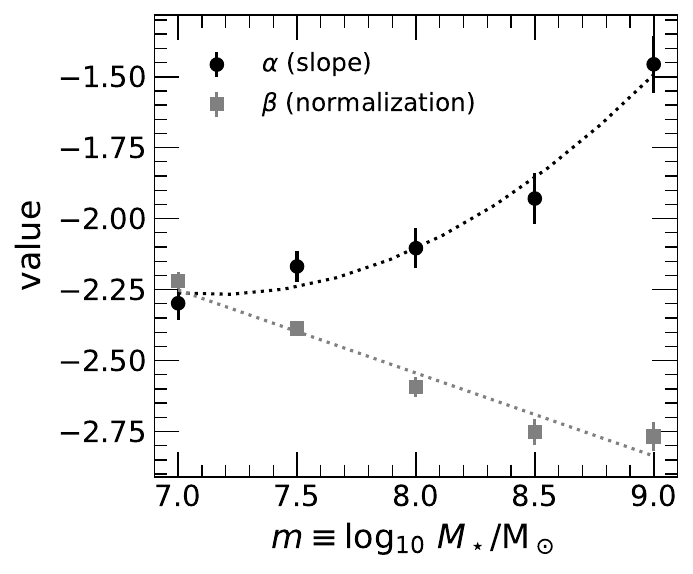}}
    \caption{A visual representation of each step in our interloper-modeling procedure. \textit{Top left}: Average number of galaxies above different minimum stellar masses, as a function of distance along host-centered (solid lines) and random (dashed lines) sightlines, at the $z = 2$ snapshot. Each minimum stellar mass limit is identified by a different color. \textit{Top right}: Average difference between number of galaxies along host-centered and random sightlines as a function of distance for different minimum stellar masses. Again, each minimum stellar mass limit is represented by a distinct colored line. \textit{Bottom left}: Number of interlopers, defined as the average value of $\langle N_{\rm{host}} - N_{\rm{random}} \rangle$ from 50 to 100 Mpc $h^{-1}$, as a function of scale factor for different minimum galaxy stellar masses. The dotted lines show the best-fit power law relationships between $\Nint$ and $a(t)$. The colors represent the same stellar mass thresholds as the top two panels. \textit{Bottom right}: The slope ($\alpha$; black) and normalization ($\beta$; gray) of the best-fit power-law relation between $\Nint$ and scale factor $a(t)$ as a function of minimum interloper stellar mass.}
    \label{fig:interloper_modeling}
\end{figure*}

Figure \ref{fig:interloper_modeling} illustrates our procedure for determining the value of $\Nint$ as a function of scale factor and minimum interloper stellar mass. The top left panel plots the average cumulative number of galaxies above five different stellar mass limits that fall within a sightline as a function of separation from a host galaxy ($\avgNhost$) or a random location ($\avgNrand$) at the $z = 2$ snapshot. As expected, we see that $\avgNhost$ experiences an initial steep increase with distance before entering the linear regime, whereas $\avgNrand$ is linear with distance throughout.

At the same snapshot and for the same set of minimum stellar mass thresholds, in the top right panel, we show the difference in average galaxy counts between host-centered and random sightlines ($\langle N_{\rm{host}} - N_{\rm{random}} \rangle$) as a function of distance. This is equivalent to taking the difference between the pairs of solid and dashed lines in the top left panel. The asymptotic value of $\langle N_{\rm{host}} - N_{\rm{random}} \rangle$ represents $\Nint$, the excess in galaxy counts due to the presence of correlated structure around massive host galaxies. At a given snapshot and mass limit, we estimate $\Nint$ by taking the mean of $\langle N_{\rm{host}} - N_{\rm{random}} \rangle$ between 50 Mpc $h^{-1}$ and 100 Mpc $h^{-1}$. This number decreases with increasing minimum stellar mass, reflecting the downward slope of the galaxy stellar mass function towards higher masses.

The bottom left panel shows the redshift dependence of $\Nint$. For the same aforementioned minimum stellar mass limits, we plot $\Nint$ at 35 snapshots between $1 < z < 3.5$. We find that the $\Nint$ decreases with scale factor, which is expected --- the universe was denser at earlier times, so a fixed radial aperture encompasses more galaxies at high-$z$ than at low-$z$. Furthermore, the relationship between $\Nint$, scale factor $a(t)$, and minimum stellar mass is well-described by a power law with a mass-dependent slope and normalization,
\begin{equation}
    \log_{10} \Nint(a, m) = \alpha(m) \log_{10}\big(a(t)\big) + \beta(m).
\end{equation}
Here, $m \equiv \log_{10} M_\star$, and $\alpha(m)$ and $\beta(m)$ are the mass-dependent power law slope and normalization, respectively. We determine the best-fit power-law relation at each stellar mass limit using a simple least-squares regression and overplot it with a dotted line.

Lastly, the bottom right panel shows the best-fitting values of $\alpha$ and $\beta$ found using the fits performed in the bottom left panel as a function of minimum stellar mass. The mass-dependence of the power law slope, $\alpha$, can be characterized with a quadratic polynomial, and the normalization of the power law, $\beta$, can be characterized by with a linear polynomial. We again perform a least-squares regression to find the best-fit $\alpha(m)$ and $\beta(m)$:
\begin{equation}
    \alpha(m) = 9.392 - 3.259 m + 0.228 m^2,~\rm{and}
\end{equation}
\begin{equation}
    \beta(m) = -0.205 - 0.292 m.
\end{equation}
With this parametrization, we can calculate $\Nint$ at any mass and redshift necessary and add it into our theoretical mass functions.

%%%%%%%%%%%%%%%%%%%%%%%%%%%%%%%%%%%%%%%%%%%%%%%%%%%%%%%%%%%%%%%%%%%%%%%%%

\section{Modeling the tidal evolution of subhalos: The effect of assumed galaxy profile on tidal track shape}
\label{app:galaxy profiles}
\renewcommand\thefigure{D\arabic{figure}}
\setcounter{figure}{0}

The galaxy tidal evolution model adopted in this work treats stellar mass loss is exponential after some characteristic turnover mass scale. This form of tidal track is based off of the empirical fits to hydrodynamic simulations from \citet{Smith2016}. However, hydrodynamic simulations suffer from a unique set of numerical problems that arise when one simulates different types of particles (e.g., dark matter and baryonic particles) at unequal masses. This use of unequal masses is not a strict requirement of hydrodynamic simulations, but is used by the vast majority of simulations. Two-body scattering preferentially transfers energy from the more massive dark matter particles to the less massive stellar particles, causing galaxies to numerically heat and expand while their halos numerically cool and contract \citep{Ludlow2019, Ludlow2021, Ludlow2023}. This introduces a potential bias into the relationship between stellar and dark matter tidal mass loss inferred from hydrodynamic simulations, as galaxies with larger stellar-to-halo size ratios lose mass more rapidly than compact systems \citep[e.g.,][]{Smith2016, Errani2022}. To investigate this effect, in this appendix, we compare the hydrodynamic simulation-derived tidal evolution model of \citet{Smith2016} against an empirical formalism for tidal stripping developed using $N$-body simulations.

%In this Appendix, we investigate the predictions of idealized theoretical models for the tidal tracks of galaxies with a range of deprojected S\'ersic profiles. We investigate systematic differences between this approach and \citet{Smith2016} and show how changes in the S\'ersic index impact tidal track shape.

We model the effect of tides on subhalos using the model developed in \citet{Errani2022}, which characterizes tidal stripping as successive truncations of the energy distributions of a subhalo's stars and dark matter. We describe subhalos and their stars using Einasto and deprojected Sérsic profiles, respectively. First, assuming spherical symmetry and velocity isotropy, we use Eddington inversion to compute the initial energy distributions of the subhalo ($dm_{\mathrm{DM}}/dE$) and the galaxy ($dm_\star/dE$). This requires defining the initial distribution functions of both the dark matter and stellar components, as well as the density of states within the initial Einasto profile. The distribution function for a given density profile $\rho(r)$ is
\begin{equation}
    f(E) \propto \frac{d}{dE} \int_E^0 \frac{d\Phi}{\sqrt{\Phi - E}} \frac{d\rho}{d\Phi},
\end{equation}
where, in our case, $\Phi(r)$ is the Einasto potential. The density of states is given by
\begin{equation}
    p(E) = \int_0^{r_{\mathrm{apo}}(E)} \sqrt{2\big(E - \Phi(r)\big)}~r^2 dr,
\end{equation}
where $r_{\mathrm{apo}}(E)$ is the maximum orbital radius accessible to a particle with energy $E$ in the potential $\Phi(r)$. From this, the energy distribution can be computed:
\begin{equation}
    \frac{dm}{dE} \propto f(E)~p(E).
\end{equation}

Then, as demonstrated in \citet{Errani2022}, tidal stripping of the subhalo gradually peels away the initial energy distribution so that eventually only the most-bound particles remain. This causes a truncation of both the dark matter and stellar energy distributions, which is well-approximated by
\begin{equation}
    \frac{dm}{d\mathcal{E}}\Big|_f = \frac{1}{1 + (0.85 \mathcal{E}/\mathcal{E}_t)^{12}} \frac{dm}{d\mathcal{E}}\Big|_i.
\end{equation}
Here, we have defined $\mathcal{E}(E) = 1 - E/\Phi_0$, where $\Phi_0$ denotes the value of the potential at $r = 0$. $\frac{dm}{d\mathcal{E}}\Big|_i$ and $\frac{dm}{d\mathcal{E}}\Big|_f$ are the initial and final energy distributions, respectively, and $\mathcal{E}_t$ is the truncation energy.

We can then compute tidal tracks (i.e., $\fstr$ as a function of $\fdm$, where $\fstr \equiv m_\star/m_{\star,0}$ and $\fdm \equiv m_{\mathrm{DM}}/m_{\mathrm{DM},0}$) by calculating $m_\star(\mathcal{E}_t)$ and $m_{\mathrm{DM}}(\mathcal{E}_t)$ and then inverting to get $m_\star(m_{\mathrm{DM}})$. 

\begin{figure*}
    \centering
    \includegraphics[width=\textwidth]{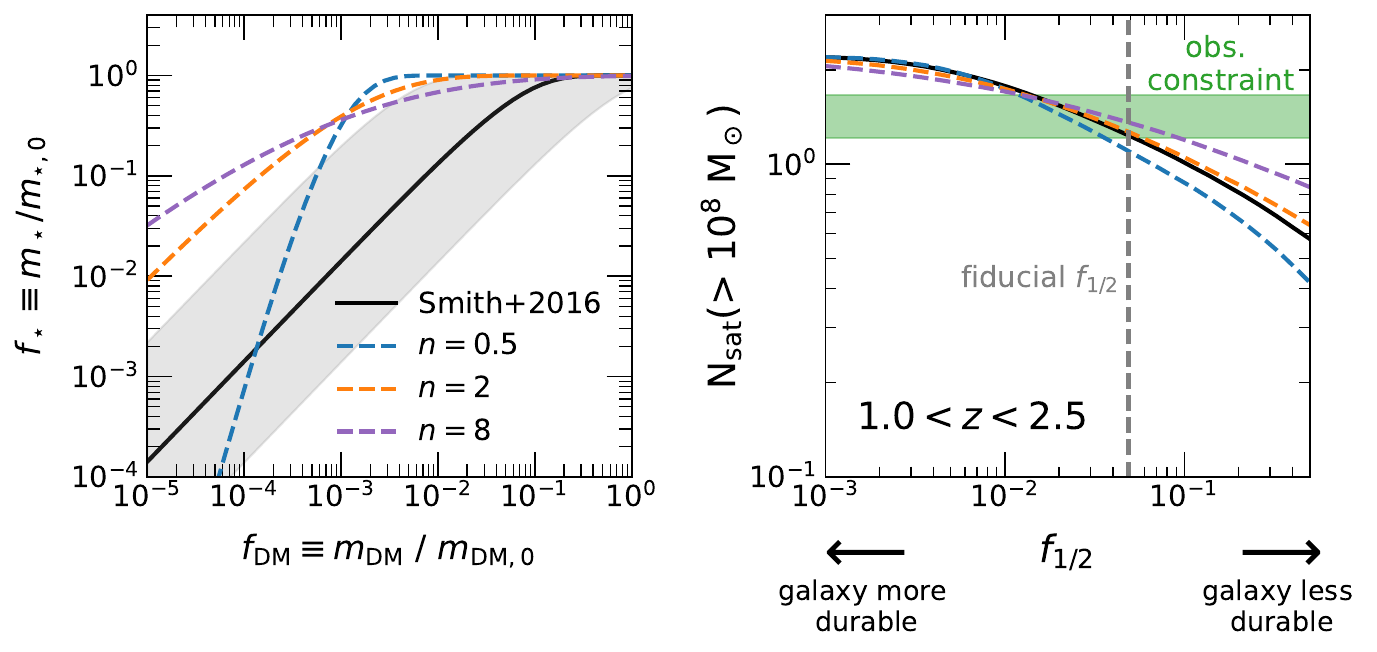}
    \caption{\textit{Left:} Tidal evolution of galaxies following deprojected S\'ersic density profiles with different indices embedded within an Einasto dark matter subhalo. The blue, orange, and purple dashed curves show the relationship between bound fraction of stellar mass $f_\star$ as a function of bound fraction of dark matter mass for galaxies with S\'ersic index $n =$ 0.5, 2, and 8, respectively, as they evolve within a static, isothermal spherical host potential. In each case, we assume the galaxy-to-halo size ratio is $r_{1/2}/\Rvir = 0.015$ \citep{Kravtsov2013}, where $r_{1/2}$ is the 3D half-mass radius of the galaxy and $\Rvir$ is its virial radius. The black solid line represents the \citet{Smith2016} tidal track used in our fiducial model. The gray shaded region indicates the range of \citet{Smith2016} tidal tracks explored by varying galaxy durability.
    \textit{Right:} The number of simulated satellites above $10^8~\Msun$ as a function of galaxy durability (parametrized by $f_{1/2}$) within the redshift range $1.0 < z < 2.5$ assuming different tidal tracks. The solid black curve shows the result from the \citet{Smith2016} tidal track parametrization and the vertical dashed gray line marks the fiducial value of $f_{1/2} = 0.0488$. The blue, orange, and purple dashed curves indicate the results assuming satellite galaxies evolve along the $n =$ 0.5, 2, and 8 deprojected S\'ersic tidal tracks shown in the lefthand panel. The green shaded band represents the satellite abundance measured from the \citet{Suess2023} observations.}
    \label{fig:sersic_tidal_tracks}
\end{figure*}

The left panel of Figure \ref{fig:sersic_tidal_tracks} shows the resulting tidal tracks for $n =$ 0.5, 2, and 8 deprojected S\'ersic galaxies (dashed blue, orange, and purple curves, respectively). These indices correspond to low, intermediate, and high central densities at a fixed $r_{1/2}$, respectively. Here, we assume the stellar-to-halo size ratio is $r_{1/2}/\Rvir = 0.015$ \citep{Kravtsov2013}, where $r_{1/2}$ is the 3D half-mass radius of the galaxy and $\Rvir$ is its virial radius. We adopt the \citet{Lima1999} approximation for the relation between the 2D half-mass radius, $r_e$, and the 3D half-mass radius, $r_{1/2}$,
\begin{equation}
    r_e = r_{1/2} \Big(1.356 - \frac{0.0293}{n} + \frac{0.0023}{n^2} \Big).
\end{equation}
This approximation has no meaningful impact on our results. The fiducial \citet{Smith2016} tidal track adopted throughout this work is plotted with a solid black line, and the range of tidal tracks we marginalized over in our main analysis is shown by the shaded gray region. 

We find that the shape of the \citet{Smith2016} tidal track matches the $n = 2$ deprojected S\'ersic tidal track, albeit with a substantial shift towards the idealized galaxies disrupting at later times. The simulated and idealized tidal tracks are nearly identical if we assume that $r_{1/2} \approx 0.1 \Rvir$, which requires the galaxy to be unphysically large. The empirical model predicts that a typically sized galaxy tends to lose mass more gradually than what is seen in hydrodynamic simulations, especially at larger $n$. This is evidenced by the fact that $f_{1/2} \approx 0.001-0.003$ in the empirical model-determined tidal tracks, over an order of magnitude smaller than the value of $f_{1/2} = 0.0488$ determined from hydrodynamic simulations. Additionally, in line with expectations, we find that at a fixed stellar-to-halo ratio, extremely centrally concentrated galaxies (S\'ersic index $n = 8$) lose mass slowly compared to their more diffuse (S\'ersic index $n = 0.5$) counterparts. These S\'ersic tidal tracks can be parametrized via
\begin{equation}
    f_\star = \Big [ 1 - \exp \big( -a (f_{\mathrm{DM}}/f_0)^b \big) \Big]^c,
\end{equation}
where $f_0$, $a$, $b$, and $c$ are functions of both $n$ and $r_{1/2}/\Rvir$. 

Furthermore, we recreate Figure \ref{fig:nsat_vs_f0} using these $n =$ 0.5, 2, and 8 tidal tracks, investigating the number of simulated satellites with stellar masses above $10^8~\Msun$ as a function of $f_{1/2}$.\footnote{In the S\'ersic tidal track parametrization, $f_{1/2}$ and $f_0$ are related as follows: $f_{1/2} = f_0 \Big [-\frac{1}{a}~\ln \big(1 - (0.5)^{1/c} \big) \Big]^{1/b}$.} This is shown in the right-hand panel of Figure \ref{fig:sersic_tidal_tracks}. The solid black curve is the result from the \citet{Smith2016} tidal track parametrization, with the vertical dashed gray line marking the fiducial value of $f_{1/2} = 0.0488$. The blue, orange, and purple dashed curves assume that satellite galaxies evolve along the $n =$ 0.5, 2, and 8 deprojected S\'ersic tidal tracks shown in the left-hand panel, respectively. The green shaded band represents the satellite abundance measured from the \citet{Suess2023} observations.

The variation of $N_{\rm{sat}}$ as a function of $f_{1/2}$ in the \citet{Smith2016} model similar to that in the S\'ersic $n=2$ model, as expected from the similarity of their tidal track shapes. For systems with higher S\'ersic indices, $N_{\rm{sat}}$ drops off more gradually with $f_{1/2}$; conversely, for systems with lower S\'ersic indices, the drop-off is steeper. This is consistent with the slope of the $n = 0.5$ tidal track being steeper, and the $n = 8$ tidal track being shallower, than the \citet{Smith2016} model after the turnover point.

We find that the $n = 0.5$, 2, and 8 curves all intersect the observational band over slightly different ranges of $f_{1/2}$. While all three yield the same lower limit of $f_{1/2} \gtrsim 0.015$, the upper limits vary: $f_{1/2} \lesssim$ 0.035, 0.055, and 0.090 for $n = 0.5$, 2, and 8, respectively. This reinforces the idea that more compact galaxies are more resistant to tidal stripping. Furthermore, while $n = 8$ is an unlikely S\'ersic index for a low-mass satellite galaxy, this also demonstrates that the choice of galaxy profile can shift the inferred bounds on $f_{1/2}$ by $\sim60\%$. Thus, accurately modeling the distribution of galaxy profiles and sizes in the simulated satellite population will be important in future work.

%TC:endignore

%%%%%%%%%%%%%%%%%%%%%%%%%%%%%%%%%%%%%%%%%%%%%%%%%%%%%%%%%%%%%%%%%%%%%%%%


\bibliography{highz_sats}{}
\bibliographystyle{aasjournal}



\end{document}